# Inductance and Mutual Inductance of Superconductor Integrated Circuit Features with Sizes Down to 120 nm. Part I

Sergey. K. Tolpygo, Evan B. Golden, Terence J. Weir, and Vladimir Bolkhovsky

Lincoln Laboratory, Massachusetts Institute of Technology, Lexington, MA, USA

***Abstract*—Data are presented on inductance of various features used in superconductor digital integrated circuits such as microstrip and stripline inductors with linewidths down to 120 nm and different combinations of ground plane layers, effect of perforations of various sizes in the ground planes and their distance to the inductors on inductance, inductance of vias of various sizes between adjacent layers and composite vias between distant superconducting layers. Effects of magnetic flux trapping in ground plane moats on coupling to nearby inductors are discussed for circuit cooling in a residual field of several configurations. Test circuits used for the measurements were fabricated in a new 150-nm node of a fully planarized process with eight niobium layers, SC2 process, developed at MIT Lincoln Laboratory for superconductor electronics and in its 250-nm node SC1, as well as in the standard fabrication process SFQ5ee. The SC2 process utilizes 193-nm photolithography in combination with plasma etching and chemical mechanical planarization of interlayer dielectrics to define inductors with linewidth down to about 100 nm on critical layers. All other processes use 248-nm photolithography. Effects of variation of process parameters on circuit inductors are discussed. The measured data are compared with the results of inductance extraction using software packages InductEx and wxLC. Part II is devoted to mutual inductance of various closely spaced features in integrated circuits, meanders, and transformers.**

## I. Introduction

RECENT advances in fabrication technology have enabled reliable fabrication of superconductor integrated circuits with feature sizes down to 250 nm [1]-[6] and increased the scale of integration from about $10^6$ Josephson junctions (JJs) per square centimeter [5] to about $10^7$ JJs per cm$^2$ [6],[7]. Reduction of feature size allows for increasing the scale of integration, but introduces new design challenges: closely spaced superconducting wires become mutually coupled and can no longer be modeled as isolated circuit inductors. While strong mutual coupling is required in some cases, e.g., for compact transformers, it is usually undesirable and should be accounted for or mitigated. Accurate knowledge of inductances and coupling of various features comprising logic and memory cells is critical for designing superconductor circuits, especially in the advanced processes.

The need for superconductor electronics (SCE) design automation (EDA) tools has stimulated development of numerical methods and inductance simulation software, initially for two-dimensional (2-D) and later for 3-D structures; see [8]-[17] and references therein. More recently, a more advanced 3-D inductance extractor, InductEx has been developed and became widely available [18]-[19]. InductEx is based on the original FastHenry approach [20]-[21] and uses finite element method for solving Maxwell and London equations for magnetic field and electric current distributions in a superconducting circuit. Recently, this approach based originally on rectangular filaments (rectangular mesh) was extended to tetrahedral volume elements (triangular mesh) [22]. A different approach was developed by M. Khapaev [13]-[17], which is based on solving London equations written for the so-called stream functions instead of electric current and minimization of the total energy functional. Using direct boundary element method, it allows numerically extract inductance, capacitance, and wave impedance of conductors with arbitrary cross sections [13]. The corresponding modeling programs are known as wxLL for inductance and wxLC for inductance, capacitance, wave impedance, and propagation speed extraction for features uniform in one directions.

A large amount of work has been done recently to characterize inductance of various structures typically encountered in integrated circuits fabricated by various fabrication processes and to compare the results with 3-D inductance extraction [23]-[26], usually obtaining a satisfactory agreement. However, the size of the superconducting features studied was relatively large, about 1 μm and larger, due to limitations of the fabrication processes used. Therefore, they do not capture and represent the peculiarities of the very dense, very large scale integration (VLSI) circuits using deep sub-micrometer feature sizes and spacing.

Previously, we reported on inductance and mutual inductance of Nb thin-film circuit inductors with linewidth down to 250 nm [27], [28] fabricated by the SFQ5ee process developed at MIT

Manuscript receipt and acceptance dates will be inserted here. This work was supported in part by IARPA via U.S. Air Force contract FA8702-15-D-0001.
*(Corresponding author: Sergey K. Tolpygo*, e-mail: sergey.tolpygo@ll.mit.edu*)*

All the authors are with Lincoln Laboratory, Massachusetts Institute of Technology, Lexington, MA 02421 USA (emails: sergey.tolpygo@ll.mit.edu, evan.golden@ll.mit.edu, weir@ll.mit.edu, bolkv@ll.mit.edu)

Lincoln Laboratory (MIT LL) [29],[30]. We obtained a good agreement, within 5% for the whole range of linewidths, between the measured inductances of straight line microsrip and stripline inductors and the extraction results using inductance extractor wxLL [13] with no fitting parameters. The agreement was somewhat worse with InductEx extraction using rectangular meshing. Agreement between the measured and extracted mutual inductances was noticeably worse in all studied cases [28].

More recently, agreement between InductEx extraction results and inductance measurements on microstip and stipline inductors fabricated by the SFQ4ee [29] and SFQ5ee [30] processes was improved by calibrating InductEx [31], i.e., by adjusting the software model settings for each circuit layer, e.g., the number of layers of filaments, filament size, the value of magnetic field penetration depth, $\lambda$, interlayer dielectric thickness, $d$, etc., to obtain the best overall agreement with the measurements [32]. The measurements were done on 12 structures involving 12 combinations of the eight niobium layers available in the SFQ5ee process stack, and with 1.2 μm linewidth, a much larger feature size than used in [27], [28]. Although an overall agreement for structures with this particular linewidth was improved to about 1% according to [32], the calibration procedure involved dozens of free fitting parameters which, unfortunately, were not constrained by independent physical measurements. For instance, for achieving the best fit, unphysically large or small values of λ could have been used for some layers and layer thicknesses were allowed to deviate from their nominal values.

In Part I of this work, we present self-inductance measurement for inductors with linewidths from 4 μm down to 100 nm, especially concentrating on the 120 nm to 250 nm linewidth range which is intended for the use in advanced fabrication process nodes described in [5] and Sec. IIA. In order to remove any ambiguity, the thicknesses of all metal and dielectric layers used in test structures and their linewidths were measured, and these actual thicknesses were used in the layer definition file (LDF) for inductance extraction. Sec. II describes circuit fabrication, test structures used, and the method of measurements as well as effects of the fabrication process parameter variations on inductance. In Sec. III and Sec. IV we present inductance data for various cases, effect of perforations in the ground planes − long moats of various width − on inductance, and characterize inductor–moat coupling at various distancesResults of the measurement are compared with inductance extraction software wxLL and wxLC [13], and InductEx [19].

Interlayers vias are the essential component of all integrated circuits but their inductance is usually not known. In Sec. V, we present via inductance measurement for sub-micrometer dimensions allowed in modern fabrication processes [5]. In Sec. VI we briefly discuss inductance of thin-film resistors used for resistive shunting of Josephson junctions. Sec. VII discusses magnetic flux trapping in ground plane moats at various protocols of field cooling of superconductor circuits and coupling of magnetic flux trapped in moats to adjacent and distant inductors as well as enhancement of this coupling by the residual field nonuniformity.

Part II of the paper will present data on mutual inductance of inductors for various linewidths and spacing, located either on the same or adjacent layers. We will also present inductance of meanders with closely spaced turns, and effect of perforations in the ground planes on various types of mutual inductors for the use in superconducting transformers required in ac-powered digital circuits.

TABLE I
TARGET THICKNESSES OF PROCESS LAYERS

| Layer/Process | SFQ5ee (nm) | SC1/SC2 (nm) | Comments |
|---|---|---|---|
| M0-M3 | 200 | n/a | Not used in SC1/SC2 |
| I0-I3 | 200 | n/a | Not used in SC1/SC2 |
| M4 | 200 | 200 | Bottom ground plane |
| R4 | n/a | 40 | Planarized resistor in SC2 |
| I4 | 200 | 260 | Between M4 and M5 |
| M5 | 135 | 135 | JJ bottom electrode[a] |
| J5 | 250 | 200 | JJ top electrode |
| I5 | 280 | 260 | Between M5 and M6[b] |
| R5 | 40 | n/a | Not planarized resistor |
| M6 | 200 | 200 | Critical layer in SC2 |
| I6 | 200 | 200 | Between M6 and M7 |
| M7 | 200 | 200 | Top ground plane in SFQ5ee |
| I7 | 200 | 200 | Chip passivation in SFQ5ee |
| M8 | 250 | 200 | Pt/Au pad metallization in SFQ5ee; Nb in SC1/SC2 |
| I8 | n/a | 200 | Between M8 and M9 |
| M9 | n/a | 200 | Nb layer in SC1 and SC2 |
| I9 | n/a | 200 | Was used as passivation |
| M10-M11 | n/a | 200 | Not used in this work |
| PAD | n/a | 250 | Pt/Au pad metallization |

[a]Thickness of Nb metal remaining after anodization of the Nb/Al bilayer
[b]Total thickness from metal to metal, including anodized surface of M5

## II. Fabrication Process, Circuits, and Measurements

### A. *Description of Fabrication Processes*

Fabrication processes SFQ5ee and SC1 developed at MIT Lincoln Laboratory (MIT LL) for superconductor electronics were described in [5],[29],[30]. They are based on patterning niobium layers, labeled Mi (i = 0,1…) by 248-nm photolithography and high density plasma (HDP) etching, with subsequent planarization of each layer. Planarization involves depositing a thick $SiO_2$ layer and its chemical-mechanical polishing to obtain a smooth planar surface and the desired thickness of the interlayer dielectric labeled as Ii. After etching contact holes in the dielectric layer Ii, the next Nb layer, Mi+1, is deposited, filling the etched contact holes, to form superconducting vias between layers Mi and Mi+1. These vias are also labeled Ii.

Due to resolution limitations of the 248-nm photolithography, the minimum linewidth, $w$ and pitch, $p=w+s$, allowed in the SC1 and SFQ5ee processes are 250 nm and 500 nm, respectively, where $s$ is spacing between parallel lines. Also, due to difficulties in filling in gaps with aspect ratio larger than ~1 by a low-temperature plasma-enhanced chemical vapor deposition (PECVD) of $SiO_2$ used in the fabrication, the minimum recommended spacing between 200-nm-thick Nb lines is 250 nm.

The main difference between the SC1 and the SFQ5ee processes is in the location of the Josephson junction layer, J5 in the process stack: it is the fifth Nb layer in the SFQ5ee and the second in the SC1 process. For design compatibility, we preserved the same layer notations in both processes. So, Nb layer count starts from M0 in the SFQ5ee and from M4 in the SC1. We use the same notations, e.g., I4, for the dielectric layers as well as for the features, contact holes, patterned in these layers and the vias formed after these etched contact holes are filled with Nb of the next metal layer.

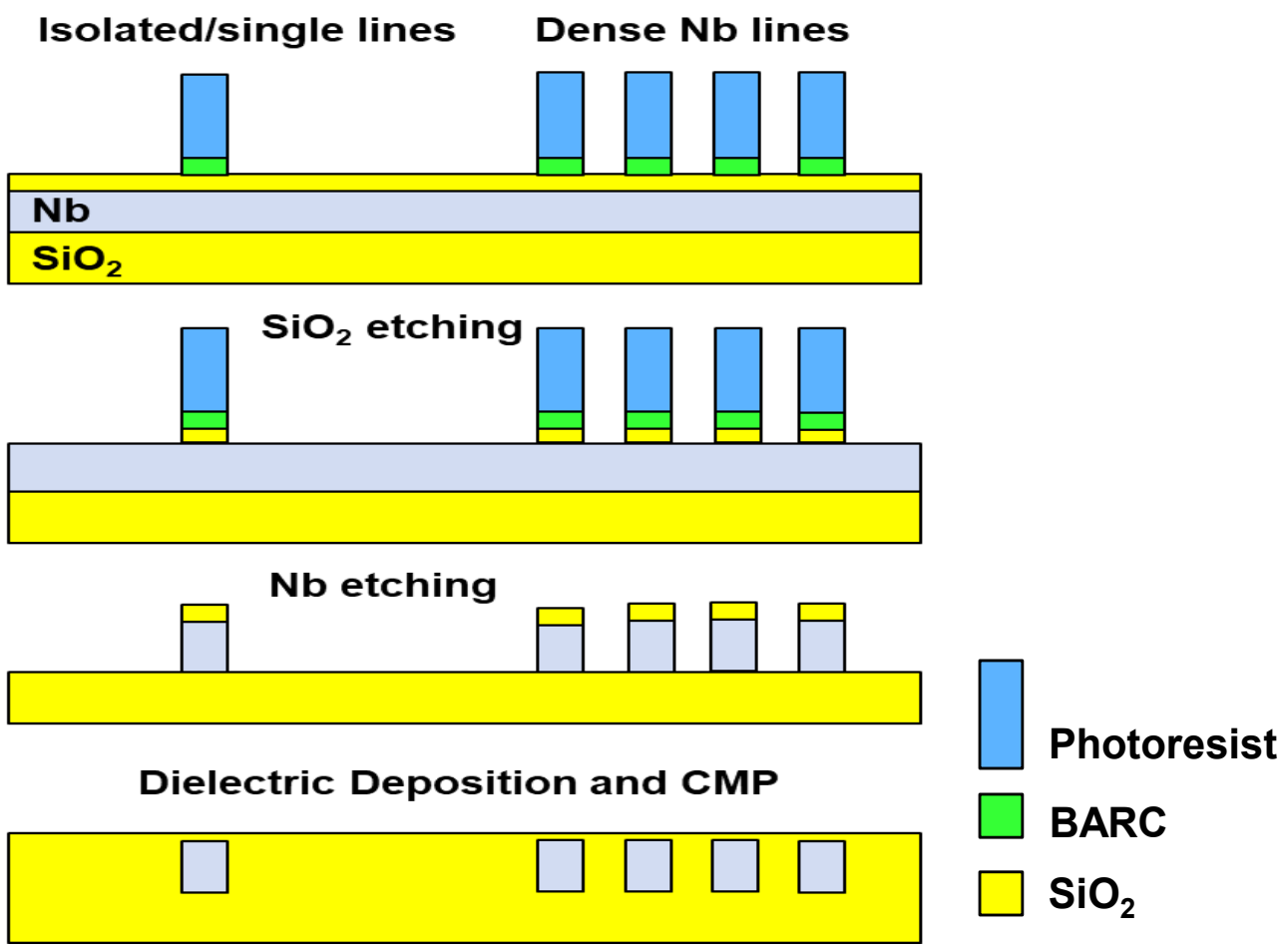

Fig. 1. Hard etch mask process used for definition of the critical Nb layers in the SC2 process employed in this work.

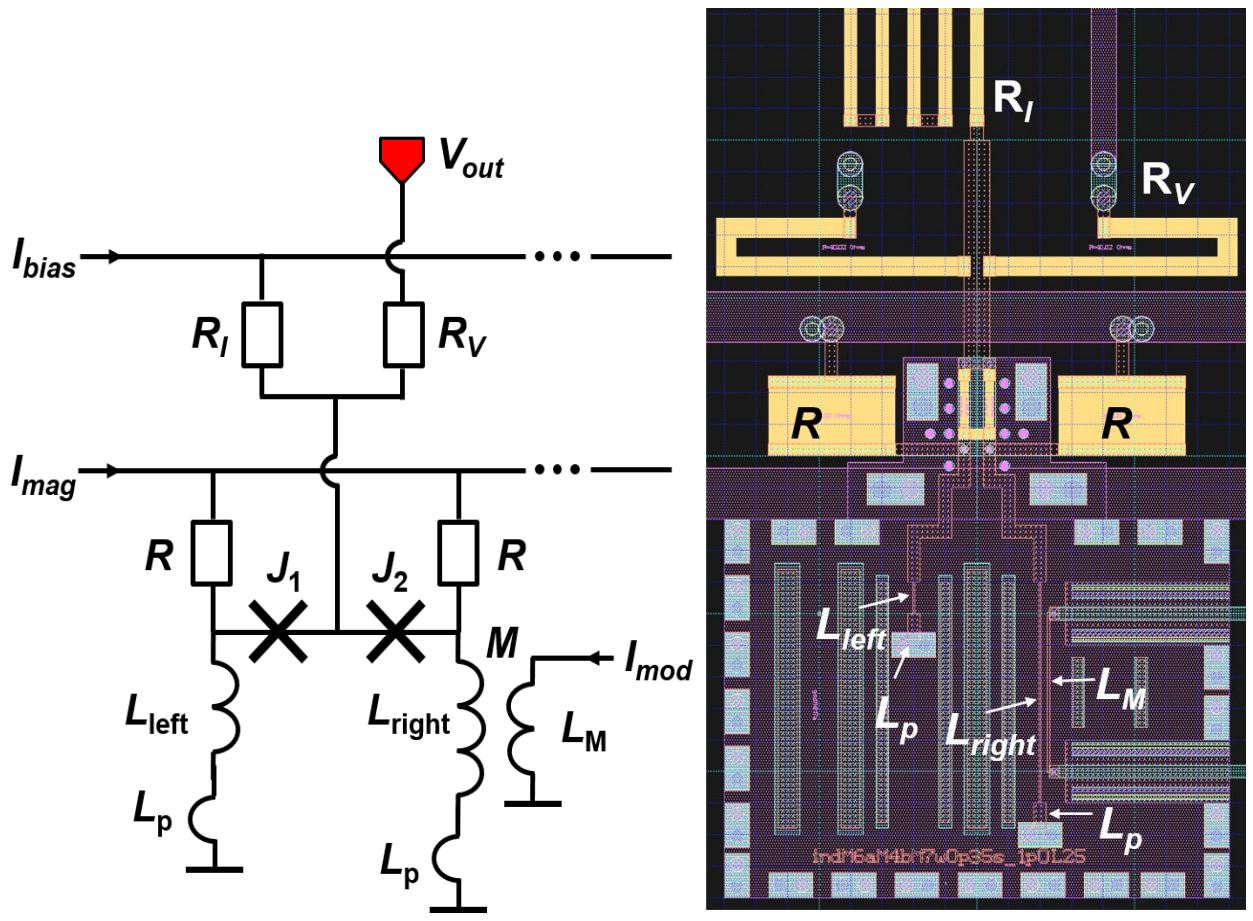

Fig. 2. Circuit diagram (left) and layout (right) of inductance measurement circuit, after [27]. It comprises $N$ SQUIDs, each formed by identical JJs $J_1$ and $J_2$, biased from a common bias rail $I_{bias}$ via bias resistors $R_I$. SQUID output voltage, $V_{out}$ is measured as a function of magnetic bias current $I_{mag}$ which is fed into both arms of each SQUID via identical resistors, $R$ dividing the current equally between $2N$ inductors. The period $\Delta I_{mag}$ of SQUID voltage modulation $V_{out}(I_{mag})$ is used to calculated the difference of the inductances of the right $L_{right}$ and left $L_{left}$ arms of the SQUID from (1). If the arms, e.g., stripline inductors, differ only by the length, this differential method eliminates parasitic contributions, $L_p$ associated with the inductors' connections to the ground and to the junctions $J_1$ and $J_2$.

To alleviate resolution limitations, the SC1 process has been recently upgraded by adding 193-nm photolithography [33] to pattern a layer of Josephson junctions, J5, and a few critical layers of inductors, e.g., layer M6, whereas keeping the rest of the fabrication process identical to the SC1. This resulted in a new node, titled SC2, allowing 150 nm minimum linewidth for critical inductors in circuits. For inductance measurements in this work, the minimum design linewidth was 120 nm and the minimum pitch for dense inductors was 370 nm.

Because the etch rate of photoresists used for 193-nm photolithography is higher, and selectivity to Nb is lower, than of the 248-photoresist used in the previous processes, a hard mask process was used [33], shown schematically in Fig. 1. The image formed in the photoresist was initially transferred onto a thin, about 70 nm, $SiO_2$ layer deposited over the Nb layer. Thus formed trilayer mask comprised $SiO_2$ layer, bottom antireflection coating (BARC), andphotoresist was then used as an etch mask capable to withstand etching of 200-nm-thick niobium features.

Target (nominal) thicknesses of all layers in the SFQ5ee and the SC1/SC2 processes used in this work are given in Table I.

### B. Self- and Mutual Inductance Measurements

Inductance and mutual inductance measurements of various features were done using a SQUID-based method [34] and an integrated circuit developed in [27] for a parasitic-free extraction of self- and mutual inductances using a differential method. The typical circuit diagram is shown in Fig. 2 and the layout of this circuit was given in [27]. Two inductors, $L_{left}$ and $L_{right}$, of the type under test form two arms of a dc SQUID. Equal currents were fed into the both arms from the common magnetic bias rail, $I_{mag}$, using two identical thin-film resistors, $R$, 10 μm long and 20 μm wide. Large dimensions of the resistors in comparison with the fabrication process spread of the resistor width with standard deviation $\sigma < 50$ nm were used to guarantee equality of their values with better than 1% accuracy. Inductors $L_{left}$ and $L_{right}$ were placed 20 μm apart so their mutual coupling could be completely neglected. They differed only by the length, being identical in all other respects.

By measuring period of the SQUID modulation, $\Delta I_{mag}$ and using flux quantization in the SQUID loop, we extracted the difference of inductances of the SQUID arms

$$L = 2N\Phi_0/\Delta I_{mag}, \tag{1}$$

where $L = |L_{right} - L_{left}|$ and $N$ is the number of SQUIDs connected to the common magnetic bias $I_{mag}$. This differential method allowed us to eliminate parasitic inductance, $L_p$ associated with connection of the inductors under test to the bare SQUID on one end and to the ground plane(s) on the other end. If $\Delta I_{mag}$ is measured with 1% accuracy and the resistors are identical within 1%, the accuracy of this method can be estimated as 1.4%.

Similarly, a mutual inductor, $L_M$ was placed parallel to the inductor $L_{right}$, and a modulation current $I_{mod}$ was fed into it from a separate current source, Fig. 2. Mutual inductance, $M$ of the two inductors determines the period of the SQUID modulation, $\Delta I_{mod}$. The circuits contained 24 dc-SQUID-based test structures per chip, grouped with $N = 6$, and allowed to extract self-inductances of 24 inductors and mutual inductances of 24 pairs of inductors.

In all cases, the differential length, $l$ of the individual inductors and the mutual running length of the parallel signal conductors in coupled inductors, $l_M$ was made much larger than the inductor width, $w$, and the spacing, $s$, between of the coupled inductors, typically $l \sim 100w$, $100s$. For instance, for all inductors with linewidth $w \leq 0.25$ μm, we used $l = l_M = 30$ μm, and longer for larger linewidths. This allowed us to neglect the edge effects of narrow, 0.25-µm, perpendicular wires feeding current into $L_M$. It is convenient to characterize long and uniform inductors by the self- and mutual inductance per unit length, $\ell = L/l$ and $m = M/l_M$, and use simulations based on an infinite line approximation, e.g., wxLL or wxLC [13], for comparing the results. For comparison with InductEx, which is a full 3D inductance extractor, the actual layout of inductors in GDSII format was used.

Hereafter, we use notations M#aM#, e.g., M6aM4, for microstips, where the first M# indicates the signal conductor layer, M6, placed above the ground plane indicated by the second M#, M4. Inverted microstrips are labeled as M#bM#, e.g., M6bM7 indicates the signal conductor on layer M6 placed below M7 ground (sky) plane. Similarly, various striplines are referred to by indicating firstly the signal conductor and then the two ground planes, e.g., M6aM4bM7 refers to a stripline with signal conductor formed on the layer M6, which is placed above M4 and below M7 ground planes. In all stripline cases, these two ground planes were connected together around their edges far away from the two inductors in Fig. 2 and at their grounded ends, using superconducting vias. This created a box-like shielding configuration.

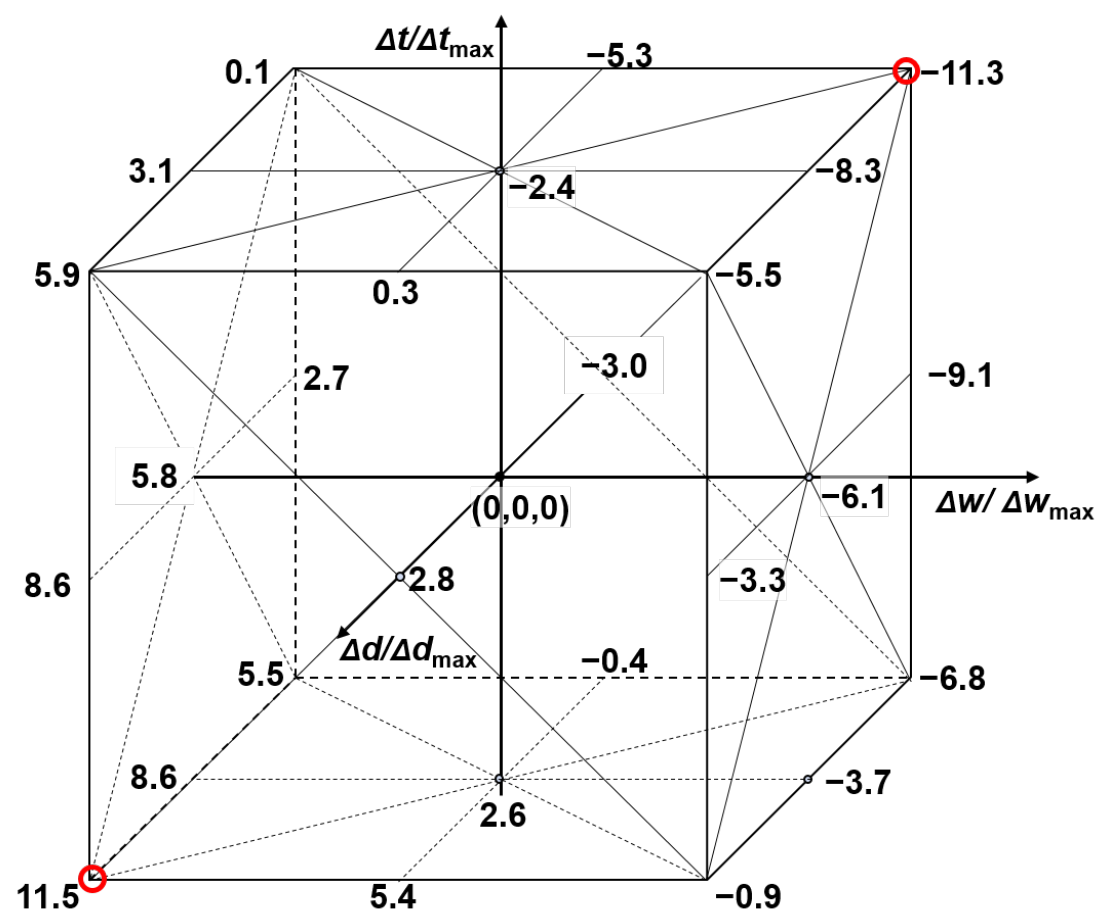

Fig. 3. Effect of process parameter deviations, $\Delta w$, $\Delta d$, and $\Delta t$ on the inductance of a M1aM0bM2 stripline with parameters: width $w = 250 \pm 25$ nm, dielectric thickness $d = 200 \pm 30$ nm, Nb film thickness $t = 200 \pm 10$ nm. The numbers on the cube surface show the percent change in inductance per unit length, $(L/L_0{-}1)$, where $L_0$ corresponds to the nominal (unperturbed) parameters. Critical corners corresponding to the largest inductance deviation in the space of process parameter deviations are marked. The simulations were done using a very accurate inductance extractor wxLL [13] and $\lambda = 90$ nm.

## C. Variations of Process Parameters and Inductance

All actual process parameters such as dielectric, $d$, and metal, $t$, thicknesses, and inductor linewidth, $w$, may deviate somewhat from the process targets (nominal parameters), varying within some ranges, $\pm\Delta w_{max}$, $\pm\Delta d_{max}$, $\pm\Delta t_{max}$. These process deviations cause deviations of inductances from their nominal (designed) values and variations across chip, chip-to-chip, and wafer-to-wafer, affecting circuit margins and circuit yield. These effects can be simulated [35] and taken into account in circuit design and margin optimization, similar to how this is done in semiconductor industry EDA tools.

It is import to distinguish between the effects of shifts in the parameter mean values, which usually have some reproducible distribution across the wafer (global variation), e.g., reflecting film deposition or CMP uniformity, and the effects of random statistical variations around the mean values. The former can be easily modeled whereas modeling the latter requires knowledge of statistical distributions of the parameters. Due to the nature of many processing tools and various process controlling feedbacks employed in the wafer fabrication, these distributions do not have to be Gaussian, and their extraction requires extensive measurements. Alternatively, it may be easier to measure the resultant inductance statistic and use it in EDA tools.

We consider here the effect of global shifts of the process parameters and discuss inductance statistics in III. As an example, let us consider a symmetric stripline M1aM0bM2 which is often used as a passive transmission line (PTL) for data and clock transmission in advanced single flux quantum (SFQ) circuits. The target parameters are: $w = 250$ nm, $d = t = 200$ nm. Variations of these parameters lay within a cube in a three-dimensional space of the normalized variables $\Delta w/\Delta w_{max}$, $\Delta d/\Delta d_{max}$, and $\Delta t/\Delta t_{max}$. From the measurements: $\Delta w_{max}$ = 25 nm, $\Delta d_{max}$ = 30 nm, and $\Delta t_{max}$ = 10 nm. The changes of the inductance caused by these deviations were simulated using inductance extraction software and shown (in percent of the expected value corresponding to the nominal parameters) on the surface of this cube in Fig. 3. Linewidth variation has the largest effect on inductance, from −6.1% to +5.8%, when all other parameters are on target. Metal thickness variation has the smallest effect on inductance, from −2.4% to +2.6%. The cube corners corresponding to the largest deviations of inductance of −11.3% and 11.5% are, respectively, $(+\Delta w_{max}, -\Delta d_{max}, +\Delta t_{max})$ and $(-\Delta w_{max}, +\Delta d_{max}, -\Delta t_{max})$. They are marked in Fig. 3 and represent the most unfavorable combinations of process parameters shifts from the target values. Realization of these extreme combinations in wafer processing is certainly possible but the least likely, whereas all other deviations cause smaller changes of the stripline inductance. For much wider inductors, $w > 1$ µm, the relative linewidth variations are small and so is their effect on inductance. Therefore, variations of inductance of wide microstips and striplines are mainly determined by dielectric thickness variations, and inductance measurements can be a sensitive method of characterizing the local dielectric thickness.

Yet another parameter that may significantly affect inductance is magnetic field penetration depth in niobium, $\lambda$, which may vary because of changes in electron mean free path caused by changes in impurity content and/or grain structure of Nb films. For the simulations in Fig. 3, we used $\lambda = 90$ nm at 4.2 K, the value that follows from prior inductance and microwave measurements on our Nb films [27],[36],[37], and [38]. Possible deviations from this value will be discussed in Sec. III.

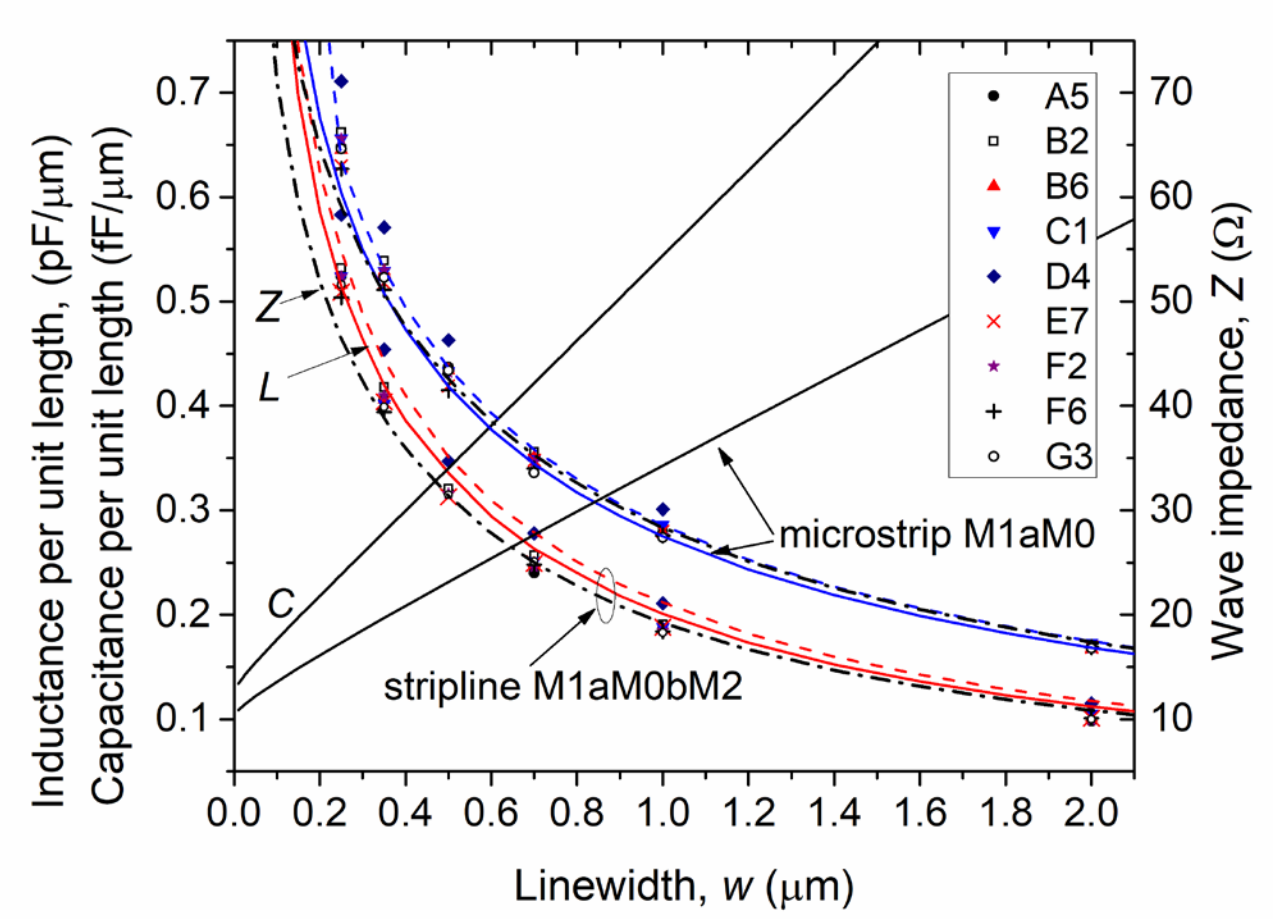

Fig. 4. Inductance per unit length and capacitance per unit length of M1aM0 microstrips and M1aM0bM2 symmetric striplines. Data for nine locations marked as A5, B2, etc. and corresponding to PCM locations on a 7x7 stepper exposure grid of the wafer are shown by various symbols. Solid red and blue curves show the simulated dependence of inductance on linewidth, using wxLC software [13] at $\lambda = 90$ nm, and dashed curves at $\lambda = 96$ nm. Black solid curves, linearly increasing with $w$, show the simulated capacitance, using a relative permittivity $\varepsilon = 4.6$ for the interlayer dielectric, PECVD $SiO_2$, as determined in [37]. Black dash-dot curves show the simulated wave impedance, $Z$ at $\lambda$ = 90 nm. Values $Z = 50.0$ Ω and 8.0 Ω are obtained at $w$ = 0.217 μm and $w$ = 2.82 μm, respectively.

## III. Inductance Results for the SFQ5ee Process

### A. *Inductance of Isolated Microstrips and Striplines*

Inductance of M1aM0bM2 striplines and M1aM0 microstips is shown in Fig. 4 as a function of linewidth of the signal conductor, for a randomly selected ran of the SFQ5ee process, SFQ523181, wafer #5 (w5). The data are shown for nine locations corresponding to the locations of process control monitor (PCM) chips with inductance test structures on 200-mm wafers. These locations are marked on a 7x7 stepper exposure grid (A,B…G; 1,2…7). The inductance per unit length, $L/l$ extracted using wxLC [13] software is shown by the solid red and blue curves, corresponding to $\lambda = 90$ nm. Also shown by solid black lines is the extracted capacitance $C/l$ per unit length, linearly increasing with $w$. The wave impedance $Z = (L/C)^{1/2}$ for both types of the transmission lines is shown by dash-dot curves. For capacitance extraction, we used a relative dielectric permittivity $\varepsilon = 4.6$ for the PECVD $SiO_2$ interlayer dielectric in the SFQ5ee process, which was obtained from resonance frequency measurements of microstrip and stripline resonators [37].

With these parameters, the value of $Z$ (in ohms) for the microstips and striplines with $w > 0.4$ μm is approximately equal to their linear inductance (in pH/μm) multiplied by 97. This means that the propagation speed of electromagnetic waves, $v_{ph}$ in wide transmission lines of these types is $c/\sqrt{9.7} = 0.32c$, where $c$ is the speed of light in vacuum. Transferring Data and Clock signals in SFQ circuits usually requires passive transmission lines (PTL) with $Z = 8.0$ Ω in order to match the typical impedance of resistively shunted Josephson junctions in PTL drivers and receivers in the SFQ5ee process. For symmetric M1aM0bM2 striplines, this value is obtained at $w$ = 2.82 μm, giving $v_{ph} = 0.324c$.

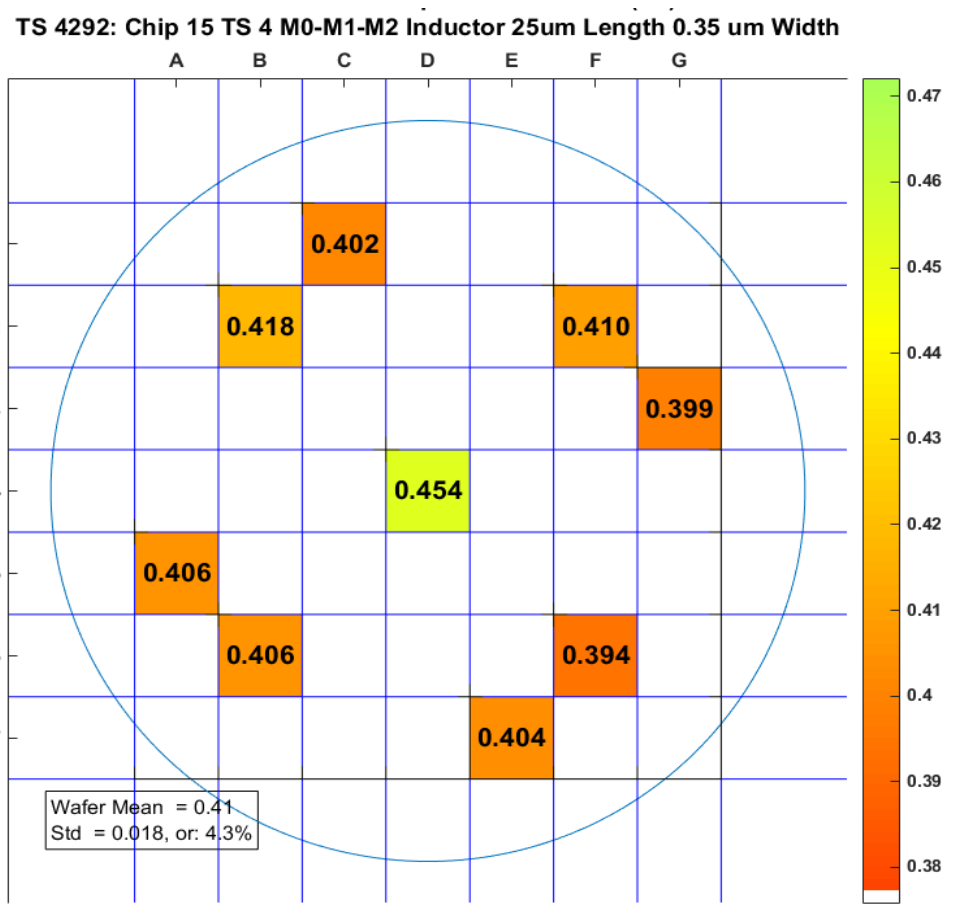

(a)

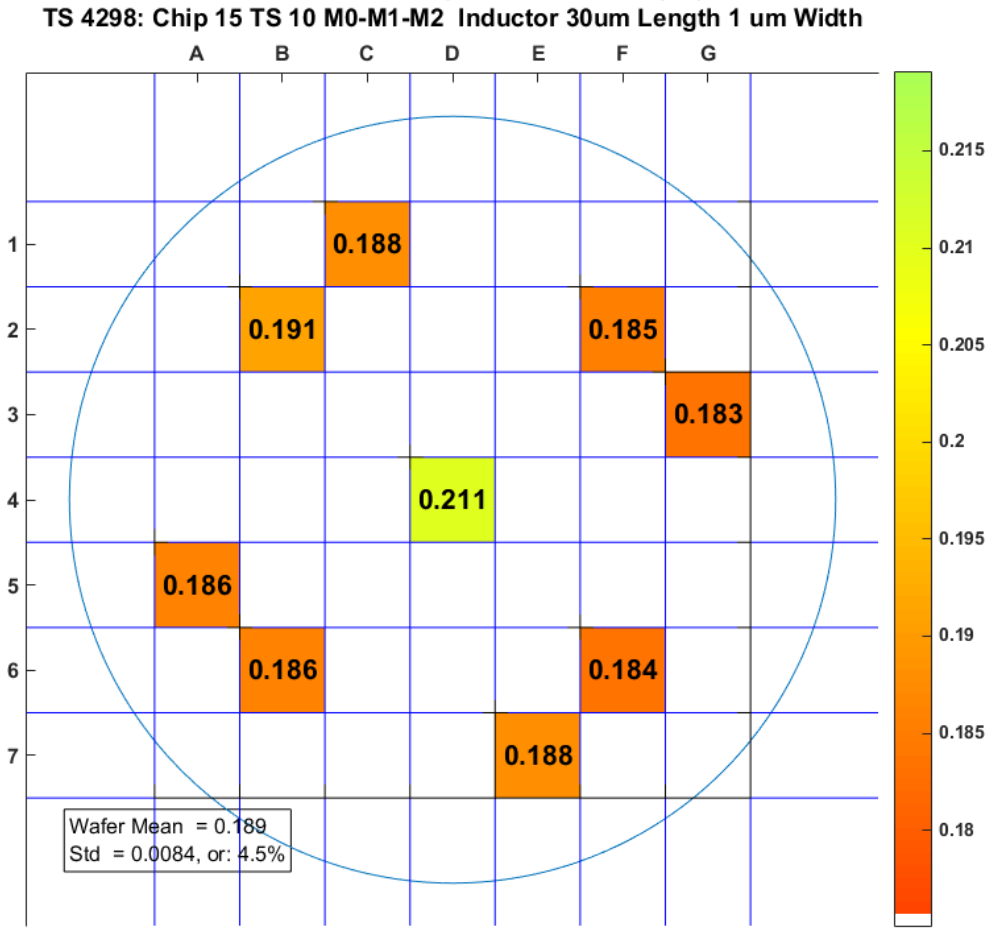

(b)

Fig. 5. Wafermaps of the linear inductance (inductance per unit length) in pH/μm for M1aM0bM2 striplines: (a) $w$=0.35 μm; (b) $w$=1.0 μm. The wafermaps correspond to nine locations of the PCM chips on 200-mm wafers in the SFQ5ee process. The wafermaps show slightly tilted distributions with approximately rotational symmetry, and a radial dependence with elevated values at the wafer center. Since the same type of wafermaps was observed for all measured linewidths up to $w = 2$ μm, this elevation is most likely caused by somewhat larger values of magnetic field penetration depth in Nb films in the central part of the wafer; see discussion in text.

Electromagnetic wave propagation significantly slows down with decreasing $w$ due to the growing contribution of kinetic inductance and fringing capacitance. It is worth mentioning in this respect that the line capacitance $C(w)$, which is linear in $w$ at large $w$, does not extrapolate to zero as could be expected from a parallel plate capacitor with infinitely thin plates. Because of its finite thickness, the signal line at $w << t$ looks like a thin metal sheet placed perpendicular to the ground planes; this sheet has a fringing capacitance which depends on $t$ but does not depend on $w$. This nonvanishing capacitance is another cause of reducing $v_{ph}$ and slowing down $Z(w)$ growth with decreasing $w$.

In Fig. 4, agreement between the measured inductance and the simulations is excellent, except for one location, D4, corresponding to the center of the wafer. A wafermap in Fig. 5 shows inductance wafemaps for M1aM0bM2 striplines with $w = 0.35$ µm, the minimum linewidth allowed by the Design Rules of the SFQ5ee process, and with $w = 1.0$ µm. The wafermaps for all other linewidths in Fig. 4 look quite similar and we do not present them to save space.

It is clear from Fig. 4 and Fig. 5 that the inductance of microstrip and striplines in the center of the wafer is, respectively, about 8% and 13% higher than at other locations of the PCM chips and that this elevation is nearly independent of the linewidth. Hence, these inductance wafermaps most likely reflect changes in the effective 'magnetic thickness' $d+2\lambda$ because, at large $w$, relative effect of linewidth deviations on inductance diminish as approximately $-\Delta w/w$ and should be negligible for $w > 0.7$ µm. The effect of $d$ and $\lambda$ deviations on wide microstrip inductors is about $\Delta L/L = (\Delta d+2\Delta\lambda)/(d+2\lambda)$. So, an 8% elevation of the inductance of M1aM0 microstrips in the central part of the wafer indicates a value of effective 'magnetic thickness' $d+2\lambda \sim 410$ nm instead of 380 nm. However, dielectric thickness measurements do not show elevated values in the central part of the wafer or a radial variation. The inductance simulations with a larger penetration depth, $\lambda = 96$ nm, are shown in Fig. 4 by the dashed curves and demonstrate a much better agreement with the data for the central die.

Rotational symmetry of many fabrication parameters and the center-to-edge radial dependence are typical for many fabrication processes due to the rotational symmetry of thin film deposition and etching tools. For instance, sheet resistance of Nb films, residual stress, tunnel barrier resistance in Nb/Al-$AlO_x$/Nb junctions, Josephson critical current density, etc., all show similar wafermaps with nearly rotational symmetry and radial dependence [5]. Therefore, $\lambda$ may also be slightly higher in the central part of the wafer where residual stress in the Nb films was found to be close to zero or slightly tensile, whereas changing to strongly compressive on going towards the edge of the wafers [5]. It is very well known [39] that $\lambda$ is expected to decrease with increasing electron density, e.g., caused by lattice compression, and to increase with increasing electron scattering rate, i.e., with increasing the film resistivity. The observed inductance wafermaps, if caused by changes in $\lambda$, agree with these trends following from the microscopic theory [39]. Additional measurements of $\lambda$ dependence on the residual stress in Nb films are required. CMP processes may skew rotational symmetry of the deposited dielectric and contribute somewhat to the skew in inductance wafermaps in Fig. 5.

Keeping inductance of various structures on target in a broad range of linewidths and for inductors formed on multiple superconducting layers is a challenging problem because of the dependence on many fabrication parameters. However, this is important for providing high yield and preserving operating margins of superconductor VLSI circuits because of their high sensitivity to global and local variations of inductors. Fig. 6 shows results of inductance tracking for M1aM0bM2 striplines on wafers fabricated in the SFQ5ee process since September 2017 and includes data on 32 tested wafers. By solid lines are shown the process target values for striplines of six different widths, corresponding to the simulated values shown in Fig. 4. The averaged data for each wafer are shown. Due to the described above inductance elevation on the central die, this die was excluded from the wafer averages.

If we characterize wafer yield based on the stripline inductance falling within a ±10% band centered on the target values, shown by dashed lines in Fig. 6, the yield is: 100% for $w =$ 2 µm; 97% for $w = 1$ µm; 91% for $w = 0.7$ µm and $w = 0.75$µm; and falls to 84% for $w = 0.35$ µm. This is why the minimum recommended linewidth for inductors on the M1 layer was set to be 0.5 µm in the Design Guide for MIT LL fabrication process SFQ5ee_v.1.3 [40], which we maintain for a number of users and for IARPA SuperTools program focused on the development of EDA tools for superconductor electronics [41]. This minimum recommended linewidth for inductors in integrated circuits is much larger that the physical resolution of the process, which for isolated wires is about 150 nm. Nevertheless, by carefully adjusting and controlling the photolithography process, much smaller linewidths than the minimum recommended can be defined and studied as described in the next section.

### *B. 200-nm Microstrips: Effect of Dielectric Thickness*

Linear microstrips with $w = 200$ nm were formed only on a single Nb layer, layer M6, which is typically used for critical inductors in the SFQ5ee process. Superconducting layers below M6 were used as various ground planes to create microstrips M6aM0, M6aM1, M6aM2, etc., differing by the dielectric thickness between the ground plane and the signal conductor, according to the SFQ5ee process. The obtained dependence of the M6 microstrip inductance on the dielectric thickness, $d$ above the ground plane is shown in Fig. 7. As easily seen, the obtained dependence cannot be described by the often-used textbook expression

$$L/l = \mu_r\mu_0(d + 2\lambda)/(w + b), \qquad (2)$$

which is valid for $w >> d$, $t$; where $\lambda = \lambda_0\coth(t/\lambda_0)$ is the thin film penetration depth, $\lambda_0$ the magnetic field penetration depth in bulk niobium, $\mu_r$ the relative magnetic permeability ($\mu_r$=1 is used hereafter), $\mu_0$ magnetic permeability of vacuum, and $b$ is a fringing factor [8], [9].

A simulated dependence of the linear inductance $L/l$ on $d$ is shown in Fig. 7 by a solid line. The simulations were done using wxLL software [13], $\lambda = 90$ nm for all Nb layers, and interlayer dielectric thicknesses from Table I. The simulated dependence provides a perfect description of the measured data for microstrips M6aM5, M6aM4, and M6aM3. However, the difference between the simulated and measured data increases progressively with $d$ in the row of M6aM2, M6aM1, and M6aM0 microstrips, reaching about 7% for the M6aM0 microstrips.

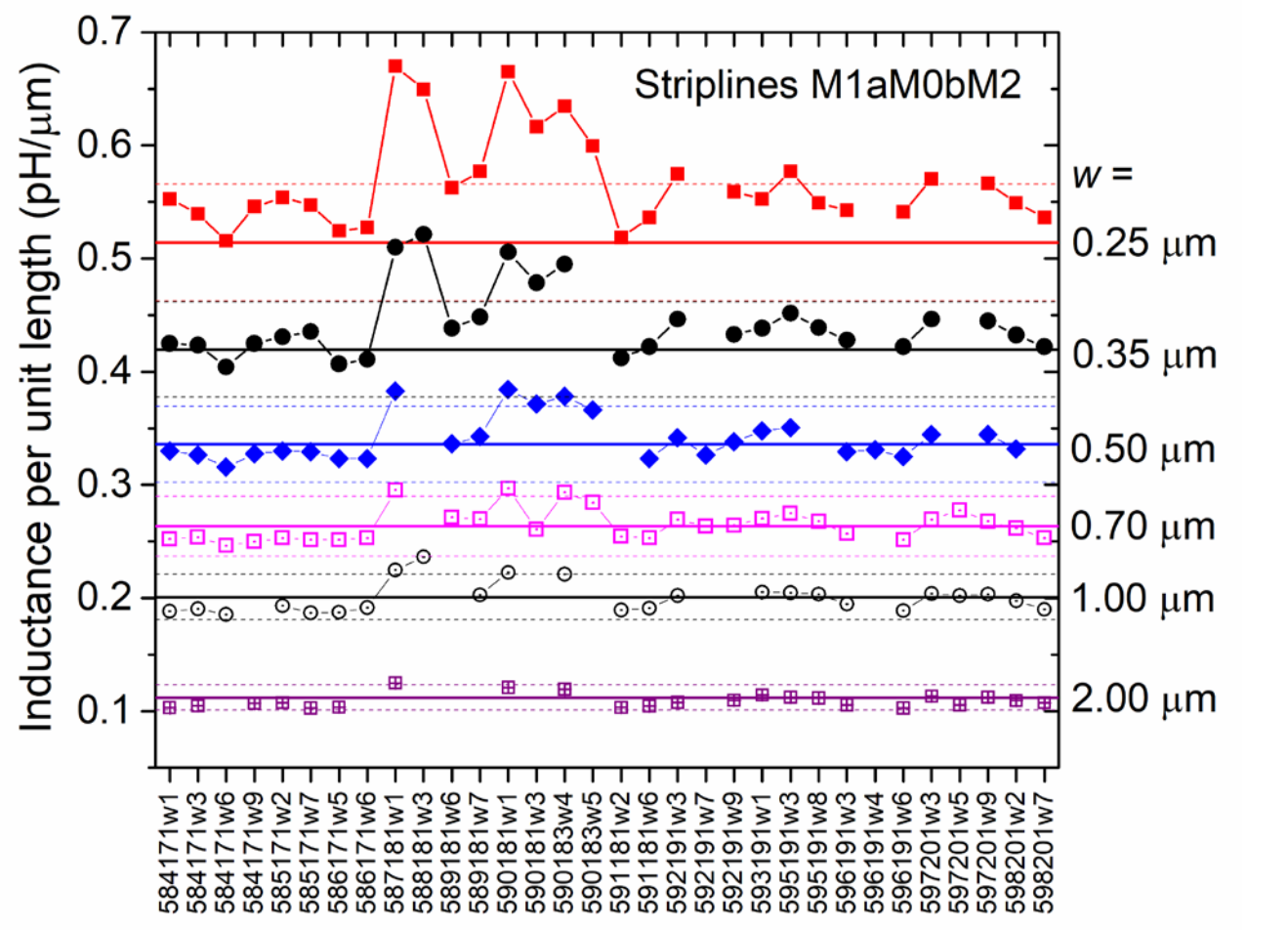

Fig. 6. Inductance tracking using 32 wafers fabricated in SFQ5ee process runs and measured since September 2017. Solid lines show the simulated, process target, values of linear inductance for each linewidth. Dashed lines show ±10% band around the target values.

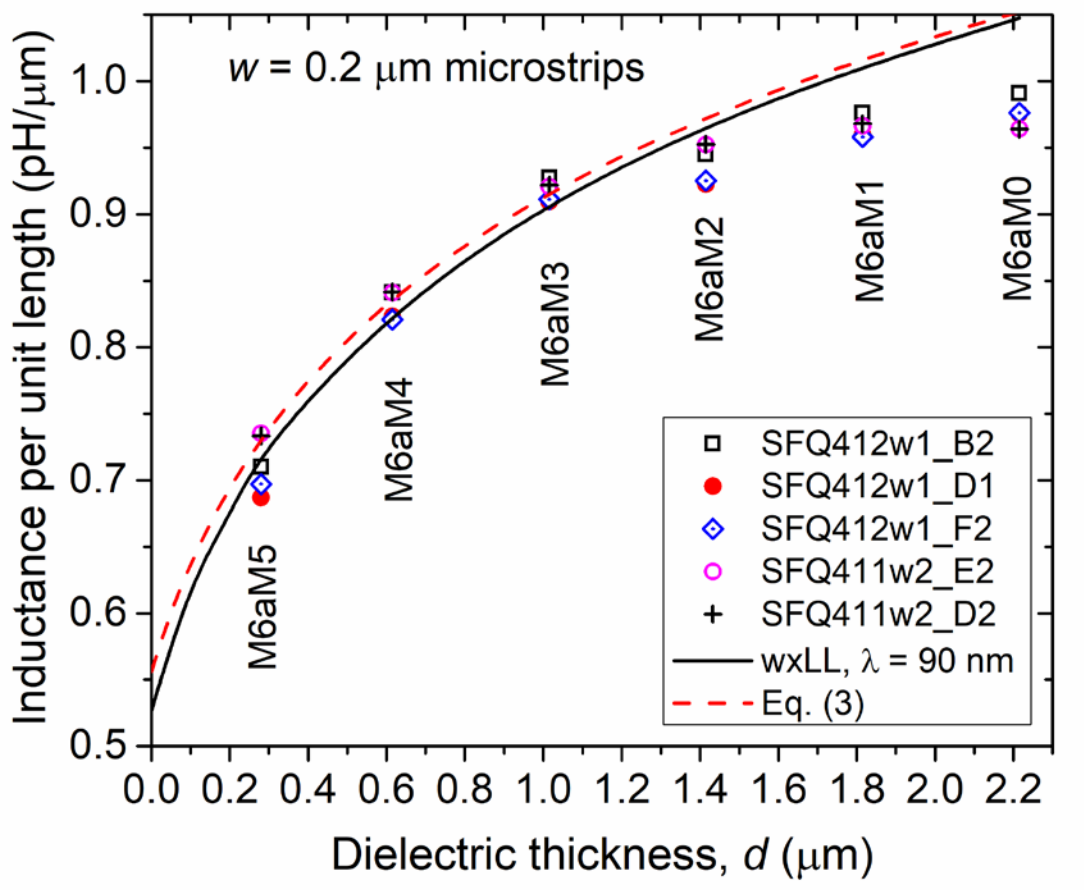

Fig. 7. Linear inductance of isolated M6 microstips placed above various ground planes in the SFQ5ee process, progressively increasing the distance between the strip and the ground plane from 0.280 μm in M6aM5 to 2.215 μm in M6aM0 according to Table I. Solid line is the simulated dependence, using wxLL software [12] and λ = 90 nm. Dashed line is (3), see III.B.

For a comparison, we also developed a simple analytical expression for the microstrip inductance. The studied microstrips have nominally a square cross section $w = t \approx 2\lambda$. Approximating them by circular wires with the same area, with radius $r = (wt/\pi)^{1/2} = 0.113$ μm, and assuming a uniform current distribution in the wire, we obtain

$$L/l = \mu_0(2\pi)^{-1}ln[2(d+r+\lambda)/r] + \mu_0(8\pi)^{-1} + \mu_0\lambda_0^2/(tw) \qquad (3)$$

The first term in (3) is due to magnetic field between the wire and the superconducting thin-film ground plane with the penetration depth $\lambda$. The second term $\mu_0/8\pi = 0.05$ pH/μm is the radius-independent part of inductance due to magnetic field inside the wire. The third term is the kinetic inductance of the wire (with uniform current distribution) per unit length, $\ell_k$. For $w = t = 200$ nm, $\ell_k = 0.2545$ pH/μm. Expression (3) is shown in Fig. 7 by a dashed line and clearly gives the same dependence of the microstrip inductance on $d$ as the numerical simulations. The values from (3) are larger only by about 3% because the last two, $d$-independent, terms in (3) are slightly overestimated as a result of assuming the uniform current distribution in the wire. The best agreement with the numerically extracted inductance is achieved if we take their sum to be 0.29 pH/μm instead of 0.3045 pH/μm calculated above.

The fabrication process of M6aM2, M6aM1, and M6aM0 microstrips requires etching away of, respectively, three, four, and five Nb layers in the stack above the ground plane and filling the volume of each removed layer with a planarized $SiO_2$ dielectric. Hence, a possible explanation of the observed deviations in their inductance would be that the actual dielectric thickness is lower than the nominal thickness $d$ used in simulations, e.g., due a dishing of the $SiO_2$ interlayer dielectric during chemical mechanical planarization (CMP) in areas with removed metal layers. However, this assumed thickness difference should be quite significant because of a weak, logarithmic, dependence of inductance on $d$, see (3). For instance, matching the measured and simulated data for M6aM0 microstrips would require the dielectric thickness to be about 1.65 μm instead of 2.215 μm nominal thickness. SEM measurements on cross sections of the inductors, made by a focused ion beam (FIB), do not show any significant deviations in the dielectric thickness. For instance, the measured $d$ between the M6 strip and M0 ground plane was 2.33 μm, very close to the nominal value.

For microstrips with dielectric filling in of $m$ etched away Nb layers $d = mt + (m+1)d_0$, where $d_0$ is the nominal thickness of a single layer of interlayer dielectric between two adjacent metal layers; $d_0 = 200$ nm for all pairs except between M5 and M6 where it is 280 nm. The maximum expected total dielectric thickness deviation is $\Delta d_{max} = \pm[m\Delta t_{max}^2 + (m+1)\Delta d_0^2]^{1/2}$, assuming that variations of the thickness of the individual layers are statistically independent. Hence, inductance variations caused by potential variations of the dielectric thickness on a chip or across the wafer should significantly *decrease* with increasing the dielectric thickness. For example, for M6aM4 microstrip inductors $\Delta d_{max} \approx 44$ nm at $d = 615$ nm. If we maintain the linewidth within ±10% of the target value $w = 200$ nm, the maximum range of inductance variations is expected to be from −8.8% to 11.0% of the nominal value, corresponding to the process corners $(+\Delta w_{max}, -\Delta d_{max}, +\Delta t_{max})$ and $(-\Delta w_{max}, +\Delta d_{max}, -\Delta t_{max})$, respectively. The largest contribution comes from the linewidth variations. For M6aM0 inductors $m = 5$ and $\Delta d_{max} = \pm 77$ nm, and the total variation range of inductance is from −6.6% to 8.3%, corresponding to the same corners of the process cube as in Fig. 3. The observed inductance variations in Fig. 7 are actually much smaller that this estimate because of a low probability of the extreme deviations of all three parameters in the same fabrication run.

## IV. Inductance Results for SC1 and SC2 Processes

### A. *Inductance in a Wide Range of Linewidths*

The SC1 process also uses 248-nm photolithography, which was centered to print 250-nm features instead of 350-nm and 500-nm features used in the SFQ5ee. The SC2 process uses 193-nm photolithography for the critical layers of inductors, only layer M6 in this particular case, which allowed us to study inductors with linewidth down to 120 nm.

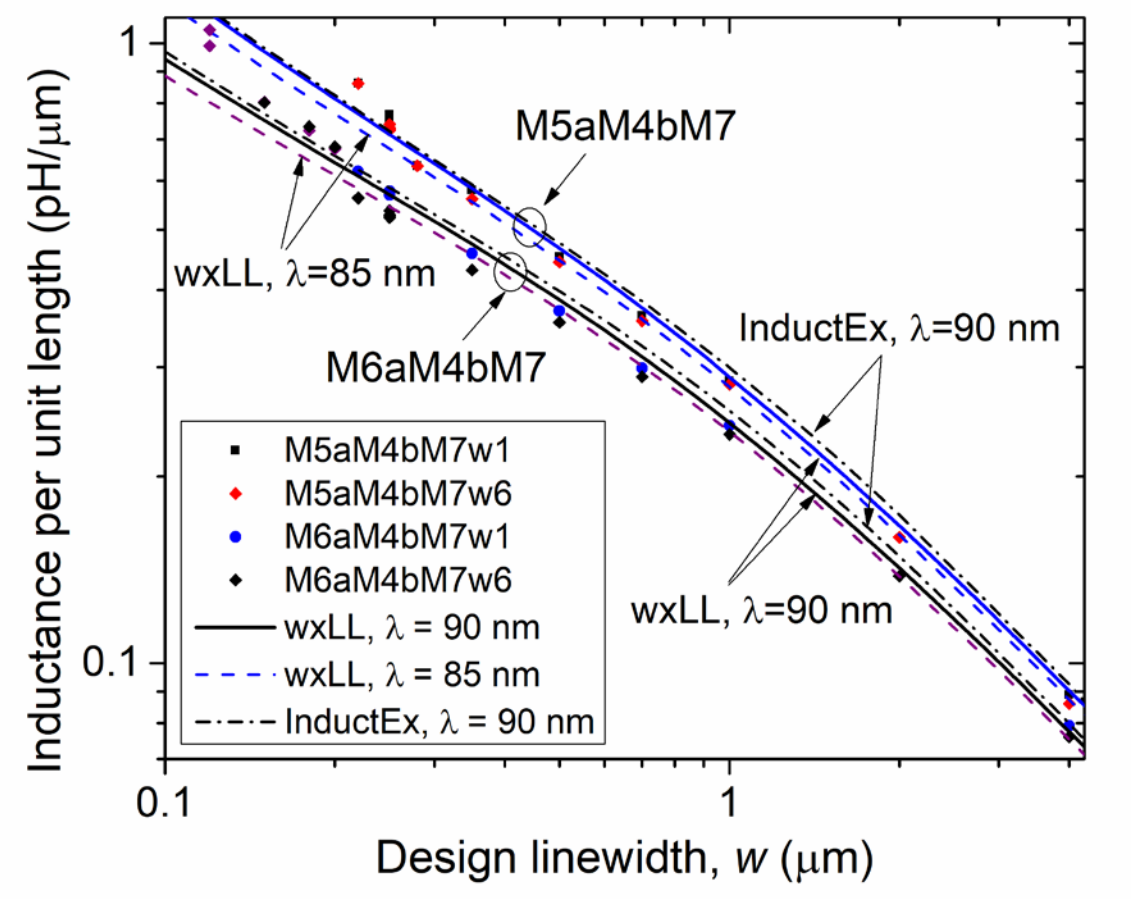

(a)

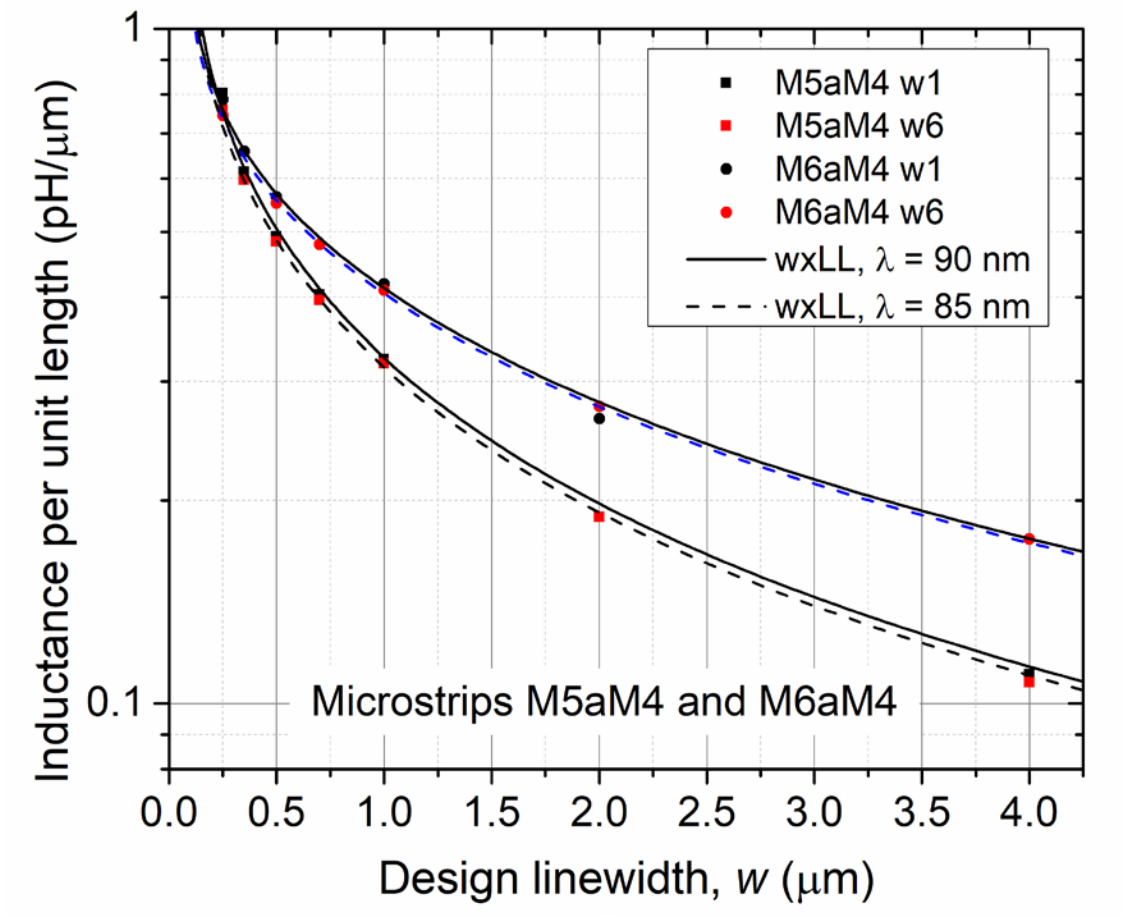

(b)

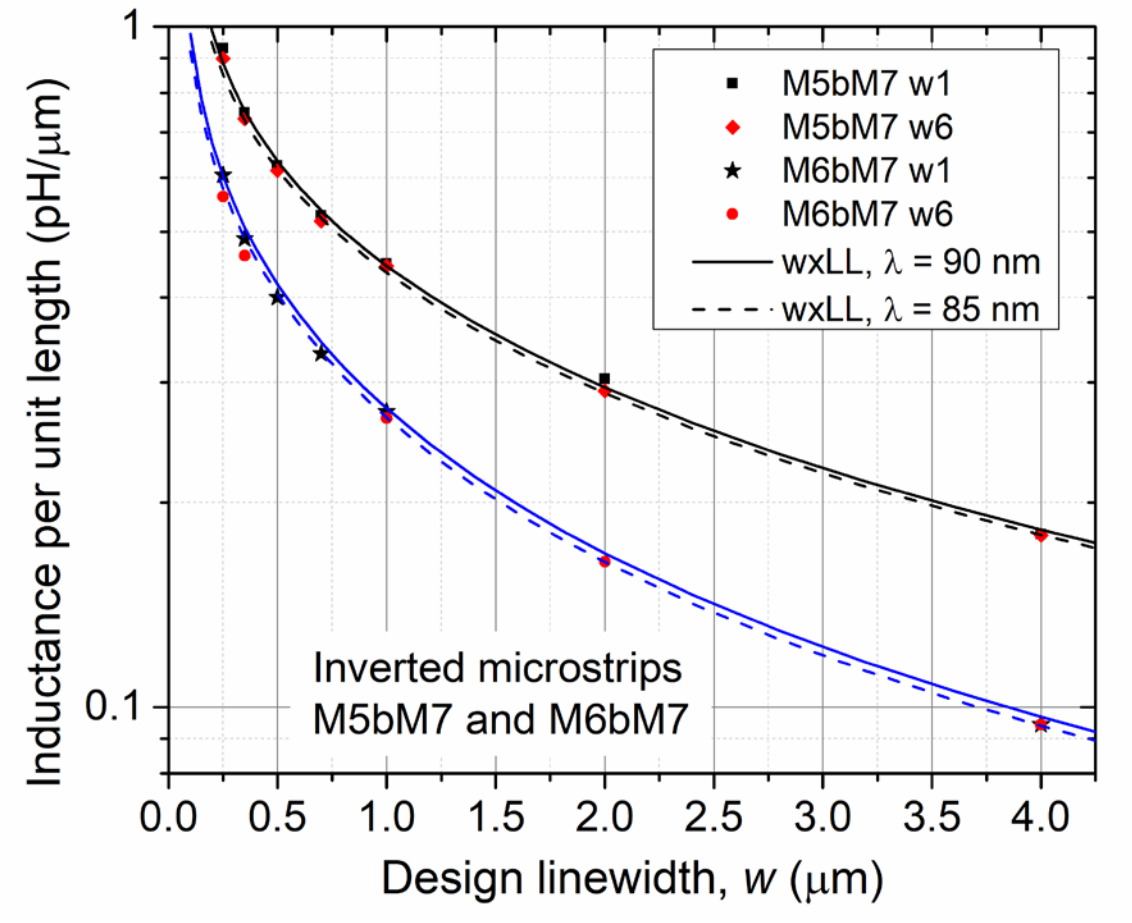

(c)

Fig. 8. Linear inductance of various isolated stripline and microstrip inductors in a wide range of linewidths in the SC2 process. M6 and M5 signal wires with linewidths, respectively, from 0.12 μm and 0.22 μm to 4 μm were patterned using, respectively, 193-nm and 248-nm photolithography in the same process. Solid and dash lines show simulations using wxLL with, respectively, $\lambda$ = 90 nm and 85 nm, and nominal layer thicknesses from Table I for the SC2 process. Dash-dot lines in (a) show simulations using InductEx with the layer definition file modified for the thicknesses in the SC2 process and $\lambda$ = 90 nm for all layers. Data are shown for two wafers, w1 and w6, which differ by the etch process used for etching niobium layer M6.

Dependences of the linear inductance of various microstrips and striplines involving layer M6 defined by 193-nm photolithography and M5 defined on the same wafers by 248-nm photolithography are shown in Fig. 8. Simulation results using wxLL are shown by solid and dash lines, and using InducEx by dash-dot lines. With nominal process parameters from Table I and λ = 90 nm, both simulation tools overestimate inductance for all *w* larger than 0.25 μm, InductEx more so than wxLL. Reducing the penetration depth to $\lambda$ = 85 nm certainly improves the agreement, indicating that the effective ‘magnetic thickness’ is smaller than the nominal. In principle, all thicknesses can be quite accurately extracted by fitting the simulations to the data in Fig. 8.

The actual linewidth on wafers, $w_w$, metal and dielectric thicknesses of inductors were measured using FIB cross sections. Multiples structures were cross sectioned to get chip-scale averages. The actual values can then be used in the layer definition files of the inductance simulators. Multiple inductor types are required for comparison to remove any ambiguity. The average data are shown in Table II for wafer #1 in Fig. 8. We can see that the actual thickness of dielectric I4 between the M4 ground plane and the signal conductor on M5 layer, $d_1$ is lower than the nominal thickness by 8%. Also, the actual thickness of the dielectric I6 between the M6 and M7 layers is lower by 10%, whereas all other thicknesses are very much on target. These lower dielectric thicknesses of I4 and I7 layers can fully explain the lower inductances of all inductors in Fig. 8 without the need to invoke a lower value of λ. Also of note is that M6 linewidth has a more significant positive bias, +35 nm, than M5 linewidth, +10 nm, which is the results of using $SiO_2$ hard etch mask in M6 processing. We will discuss the effect of linewidth bias on the narrow lines below.

## B. *Stripline Inductors with Linewidths Down to 120 nm*

Wafermaps of inductance of M6aM4bM7 stripline inductors with linewidth from 200 nm down to 120 nm are shown in Fig. 9. Stripline inductors with $w \geq$ 180-nm linewidth have less than 1% inductance standard deviation on 5-mm chips. Wafermaps show the larger inductance values in the central part of the wafer with center-to-edge variation of ~ 9%, similar to Fig. 5, but with different locations of the PCM chips in the SC2 process than in SFQ5ee process. 150-nm stripline inductors have less than 3% on-chip spreads of inductance, which is acceptable for the use in VLSI integrated circuits. On-chip inductance variation of 120-nm inductors is relatively big and its cause needs further investigation.

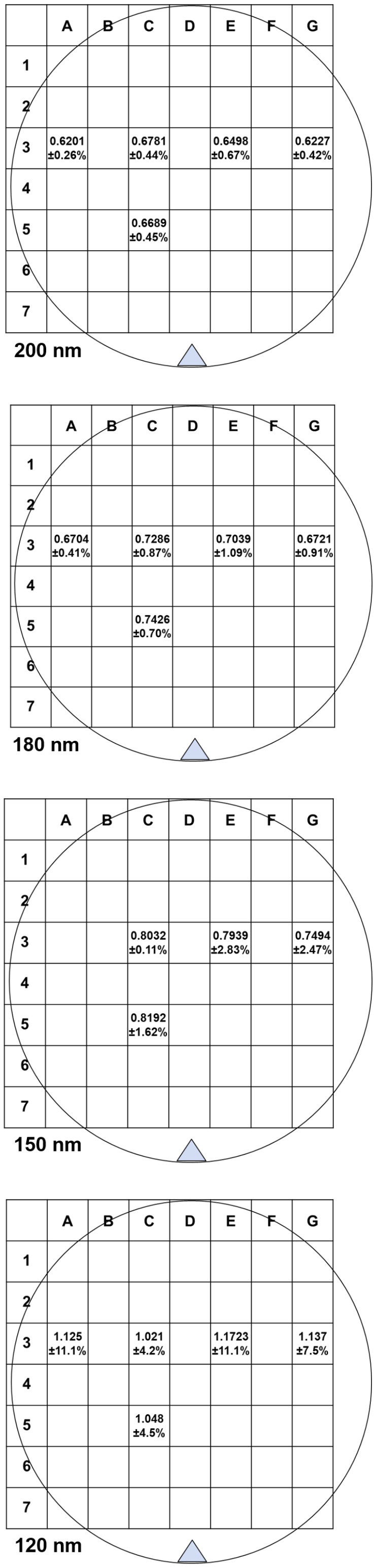

Fig. 9. Wafermaps of the linear inductance of isolated striplines M6aM4bM7 with design linewidth, $w$ from 120 nm to 200 nm. The data are for w1. Triangle at the bottom indicate positon of the wafer notch orienting wafers in processing tools. For $w \geq 150$ nm, the standard deviation of inductance on chip is less than 3%, which is acceptable for superconductor VLSI circuits. Center-to-edge inductance variation is similar to that in Fig. 5, although the SC2 uses different locations of the PCM chips than the SFQ5ee process.

TABLE II
ACTUAL THICKNESSES OF PROCESS LAYERS FROM FIB CROSS SECTIONS

| Induct/Thickness | $d_1$ nominal (nm)[a] | $d_1$ actual (nm) | $d_2$ nominal (nm)[b] | $d_2$ actual (nm) |
|---|---|---|---|---|
| M5aM4bM7 | 260 | 239 (−8%)[c] | 660 | 668 |
| M6aM4bM7 | 655 | 664 (+1%) | 200 | 180 |
| Metal/Thickness, Linewidth | Nominal (nm) | Actual (nm) | $w$, nominal (nm) | $w_w$, actual (nm) |
| M4 | 200 | 196 (−2%) | | |
| M5 | 135 | 138 (+2%) | 350 | 360 |
| M6 | 200 | 202 (+1%) | 350 | 386 |
| M7 | 200 | 204 (+2%) | | |

[a]Dielectric thickness between the M4 ground plane and the signal layer
[b]Dielectric thickness between the signal layer and the M7 top ground plane
[c]Percent deviation from the nominal value

Another important test for application of narrow-line inductors in VLSI circuits is to verify that there is no difference between horizontally and vertically oriented inductors because they are randomly mixed in superconductor integrated circuits. A difference between parameters of vertically and horizontally oriented devices, e.g., resistors and inductors, may indicate astigmatism of the projection optics, causing linewidth difference, or/and texture in the film growth (e.g., for thin-film resistors). Fig. 10 presents data for the horizontal and vertical inductors marked, respectively, as _h and _v for seven different locations of the test chip on the wafer, which do not show statistical difference. For instance, for $w$ = 180 nm striplines, the differ-ence between the mean values for vertical and horizontal inductors on-chip is within ±1.4%; a ten-chip wafer average of this difference is −0.27%.

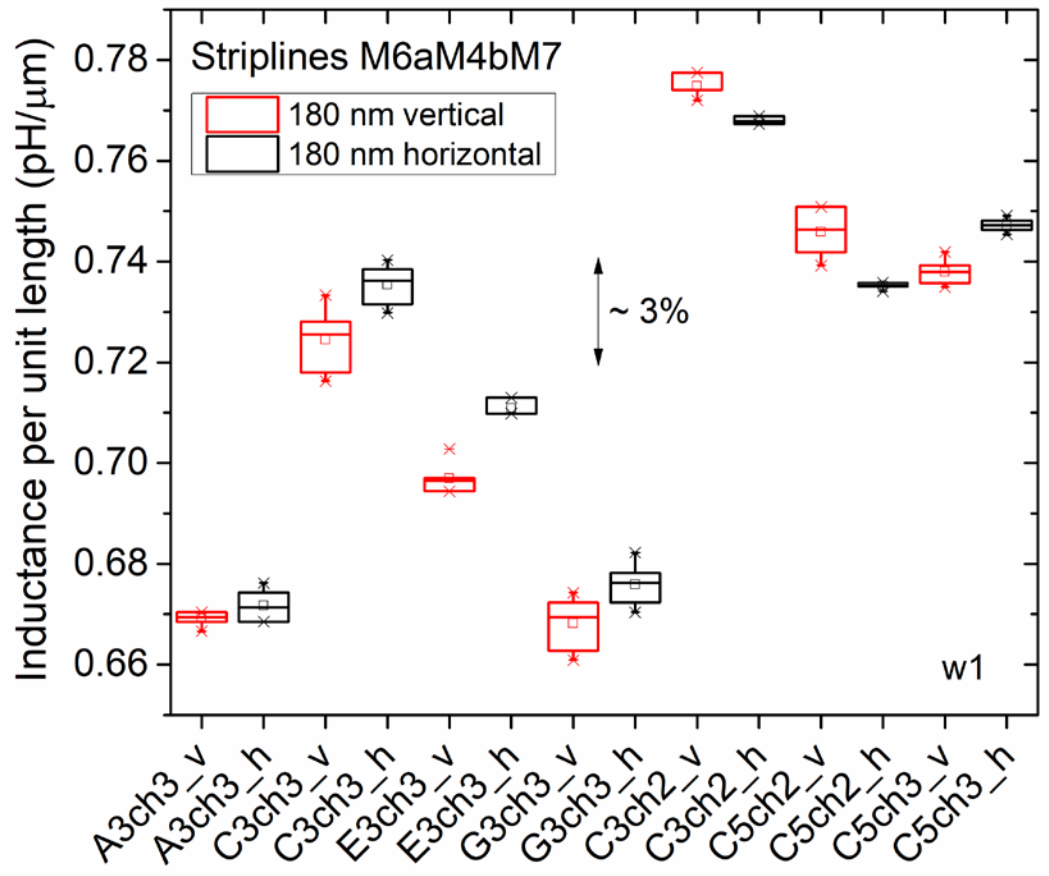

Fig. 10. Linear inductance of vertical, _v, and horizontal, _h, isolated stripline inductors with design linewidth $w$ = 180 nm on two chips, ch2 and ch3, at various locations A3, C3, etc., on the wafer. The height of the statistic boxes indicates the standard deviation, 1σ. Also indicated are the mean and the average values as well as the full range of data variation. The data show that there is no statistical difference between vertical and horizontal inductors.

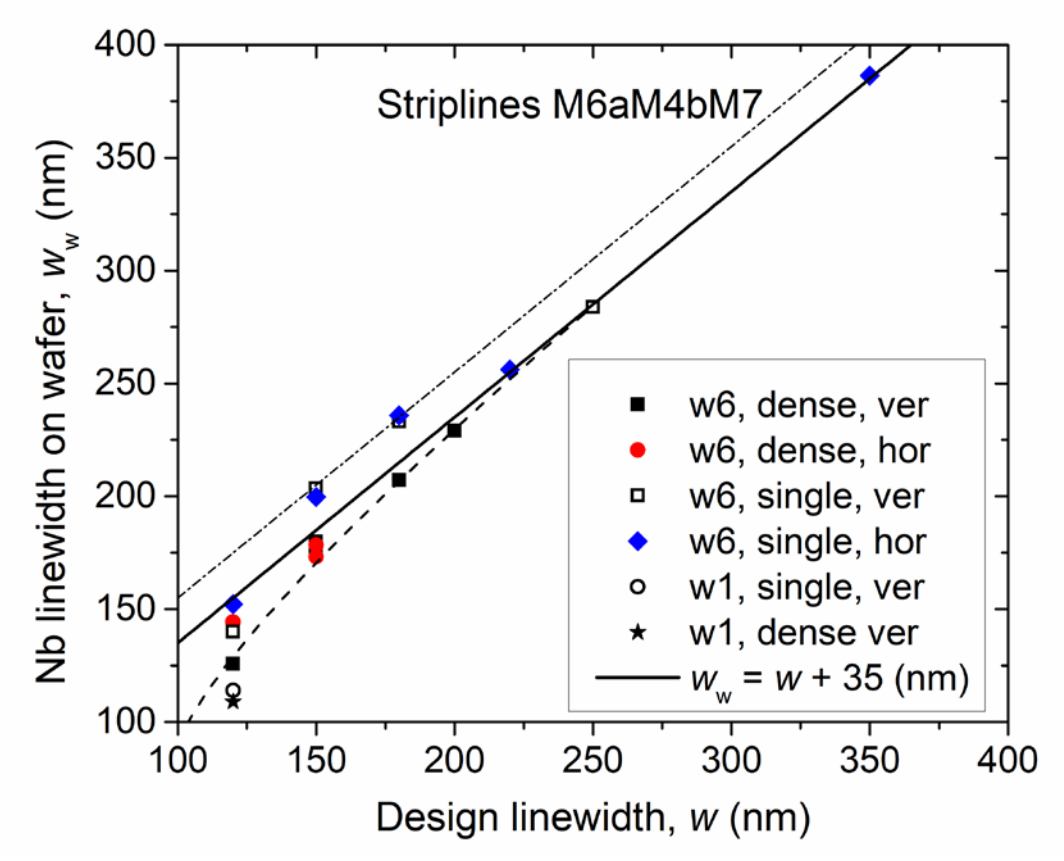

Fig. 11. The actual linewidth of Nb striplines, $w_w$ measured using SEM *vs* their design linewidth, $w$. The data are shown for dense lines spaced by 250 nm and single, isolated, lines, both vertical and horizontal. A solid line shows constant positive process bias of 35 nm. A dash line shows a potential roll down of the linewidth of isolated lines. Please note a difference between the linewidth of the dense and isolated lines. Also, the process bias for w1 is smaller, and slightly negative, than for w6 due to an intentional difference in the etch process.

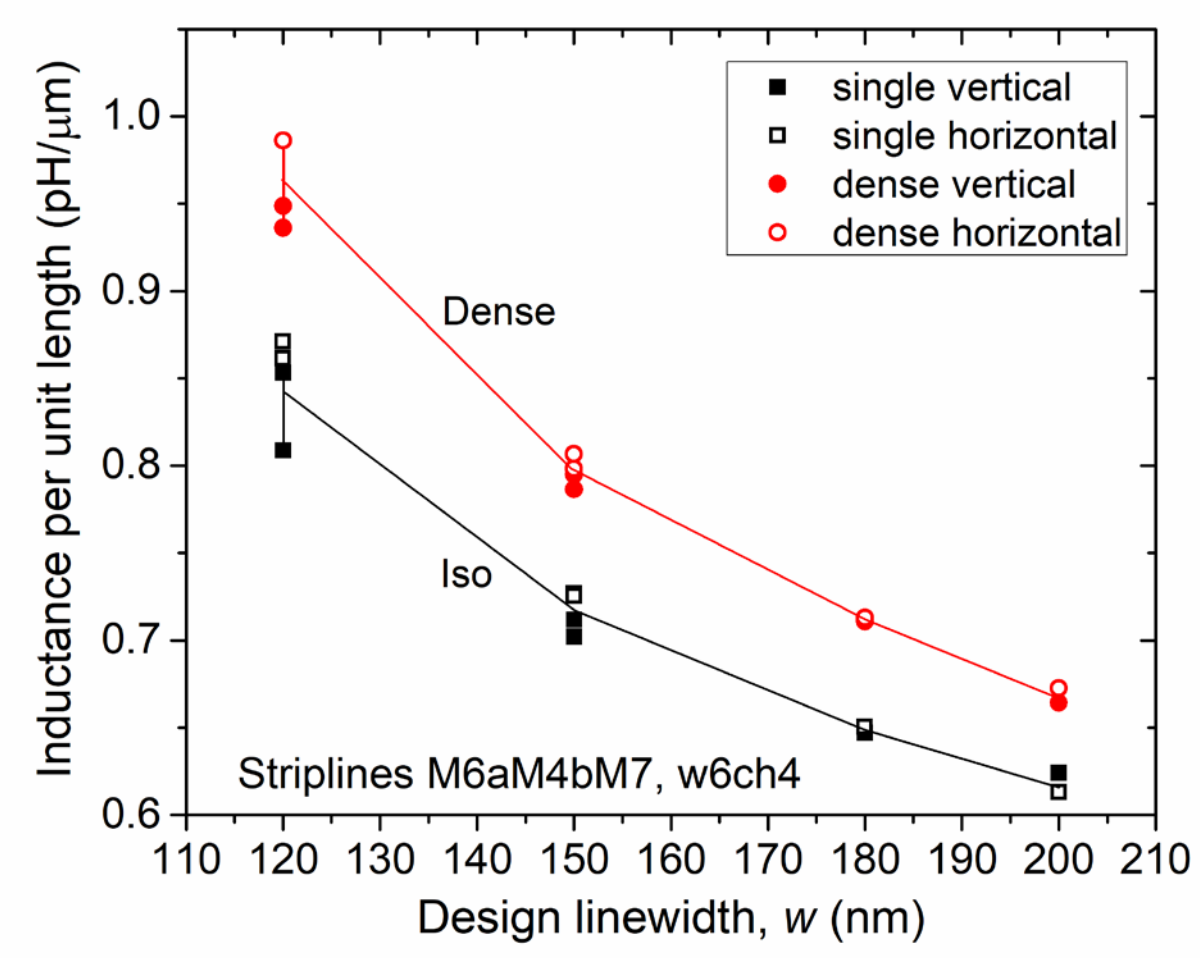

Fig. 12. Linear inductance of single (isolated) and dense stripline inductors M6aM4bM7 with different design linewidth, $w$ and two orientations: vertical and horizontal. Lines connecting data points are to guide the eye. A larger inductance of dense lines can result from a smaller physical linewidth, $w_w$ of the dense lines than of the single lines at the same design linewidth, $w$.

These data are consistent with linewidth measurements of the inductors, using SEM imaging of their cross sections done by FIB, shown in Fig. 11. The linewidth measurements also show no difference between the vertical and horizontal inductors. However, the data show a systematic difference between linewidths of isolated and dense inductors. The term 'dense' here applies to multiple, typically five, lines at the minimum spacing allowed in the process, 250 nm in this case, whereas isolated, single, inductors do not have any neighbors within a few micrometers from them. This so-called iso-dense bias is a very well-known lithographic phenomenon [42]. It can also be created by a faster etch rate in the dense areas of the photoresist mask than in the sparse areas caused by various plasma microloading and resist charging effects; see, e.g., [43], [44]. For $w$ < 200 nm, the iso-dense difference is about 20 nm, i.e., the single lines are wider than dense by about 20 nm. The difference diminishes with increasing spacing above ~ 350 nm.

Inductance of the dense and single, both vertical and horizontal, lines marked, respectively, d_v, d_h, s_v, and s_h is compared in Fig. 12 for several linewidths. It is clear that dense inductors have substantially larger inductance values, for all linewidth studied. E.g., inductance of the dense 120-nm lines on w1 and w6, is larger by, respectively, 25% and 11% than inductance of the single lines. The difference can result from a smaller actual linewidth of the dense lines than of the isolated ones, see Fig. 11, and potentially larger contamination of niobium in dense structures, causing an increase in the penetration depth. The difference between w1 and w6 can also be explained by the difference in the etching process, resulting in the overall smaller linewidth on w1 than on w6.

In Fig. 13 we plotted inductance of the single and dense inductors, both vertical and horizontal, as a function of their actual line width, $w_w$ measured using FIB cross sections. In this representation, a clear separation between single and dense lines goes away and all inductors align more or less along the same curve shown by a red dash line. For comparison, by black solid line we show inductance simulations, using wxLL, with the actual metal and dielectric thicknesses measured on this wafer.

It is clear from Fig. 13 that using the actual linewidth for wide inductors with design $w$ > 250 nm improves agreement between the measured and the simulated values, making it better than 1% for all the widths up to 4 μm. However, the linear inductance values for the physical linewidths smaller than about 250 nm are noticeably larger than those following from numerical simulations for these physical linewidths, shown by the solid line in Fig. 13. This difference can be explained by growing, with decreasing the linewidth to below the film thickness, magnetic penetration depth in Nb, from 90 nm in wide strips to ~ 105 – 110 nm in the narrowest ones. Such an increase is known to occur in planar Nb film with decreasing their thickness and caused by decreasing electron mean free path [38]. Another causes could be Nb surface oxidation or hydrogen contamination in processing; see [45], [46] and references therein. Since Nb contamination mainly occurs through the sidewalls of the etched lines, the degree of contamination grows with decreasing the linewidth because the strip surface to volume ratio increases as $1/w$. Hence, at the same processing conditions, narrower lines will always absorb more impurities and have shorter electron mean free path than wider lines.

The presence of hydrogen contamination can be detected using electromigration, similar to [46], to drive hydrogen out of the inductor and into niobium ground plane, and then observe electromigration-induced changes in the stipline inductance. We leave these experiments for a different publication.

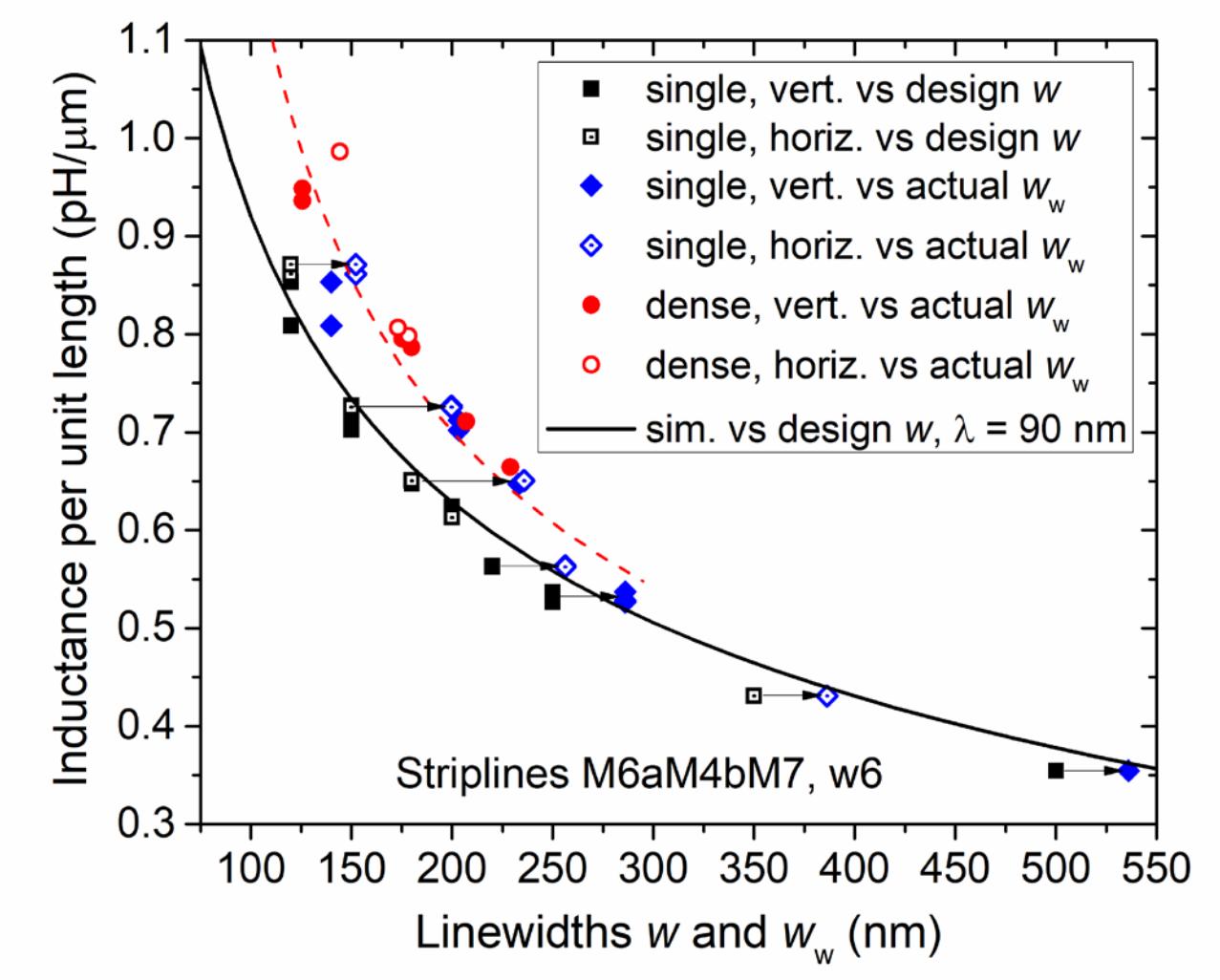

Fig. 13. Linear inductance of single (isolated) and dense stripline inductors M6aM4bM7 with different design linewidth, $w$ and two orientations replotted as a function of their physical linewidth, $w_w$ measured in SEM using FIB cross sections of the inductors. The same data for the single inductors are also shown as the function of their design width. Arrows show the shift of the data points when the design linewidth is replaced by the actual linewidth. Black solid line shows the simulated inductance using wxLL with the actual dielectric and metal thicknesses measured on this wafer. Red dash line corresponds to a 35 nm shift of the origin of the x-axis for the simulated curve. A good agreement with the data in this case suggests that that an electric (superconducting) width of the wires is about 35 nm smaller than the physical width, e.g., due to the presence of a thin nonsuperconducting sheath on the sidewalls of the etched Nb lines.

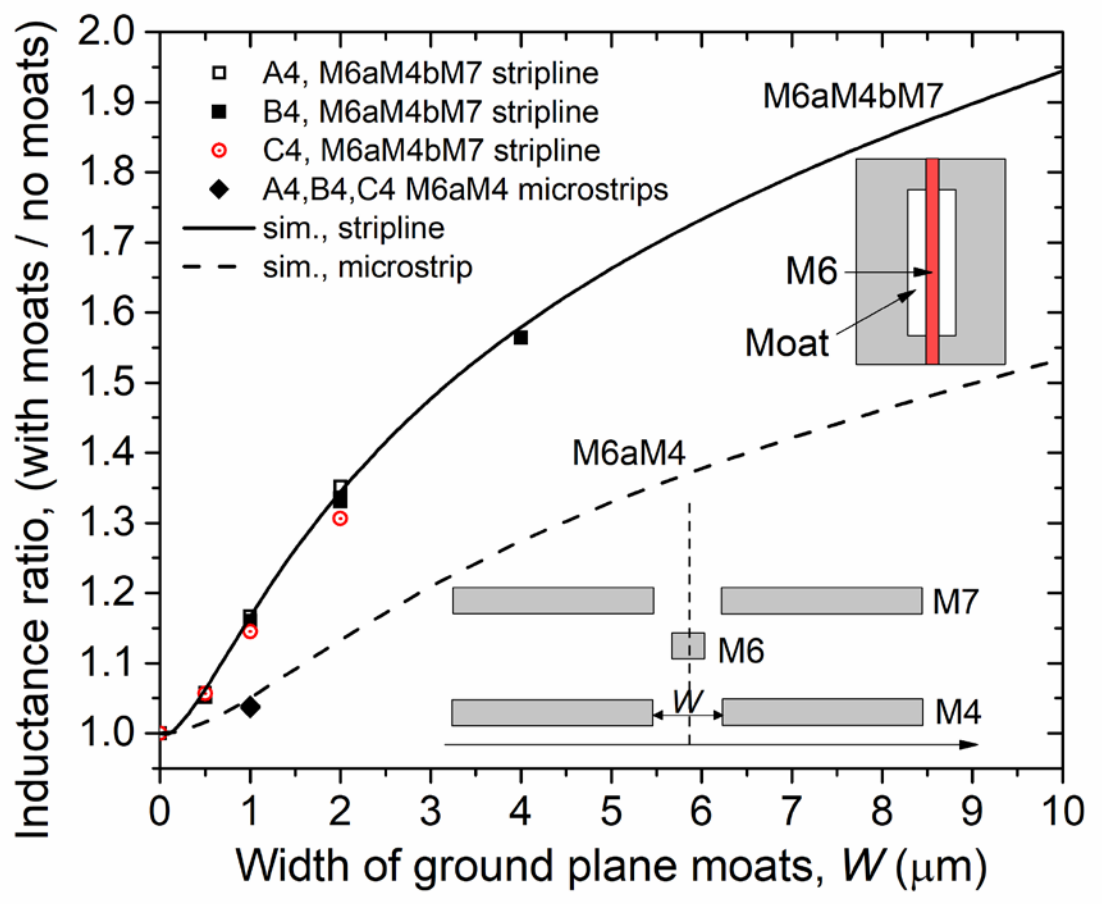

Fig. 14. Relative change, $L/L(0)$, in inductance of isolated stripline M6aM4bM7 and microstrip M6aM4 inductors, introduced by symmetrically placed moat(s) in the ground plane(s) M4, for microstips, and congruently in M7 for striplines, as a function of the moat(s) widths, $W$. The linewidths of inductors is $w = 0.25$ μm. The length of the moats is 30 μm, much larger than $w$ and $W$. The inductor length is 35 μm, longer than the moats, but the contribution of its ends running between unperforated ground planes is deducted in the differential method used in the measurements. Data points correspond to three locations, A4, B4, and C4, on the wafer fabricated by the SFQ5ee process. The measured inductance was normalized to $L(0)$, the inductance value at $W = 0$ corresponding to the unperforated ground planes. The simulated dependences are shown by solid and dashed line, respectively, for the striplines and microstrips, using wxLL software [13] with layer parameters for the SFQ5ee process in Table I and $\lambda = 90$ nm. A cross section and top view of the inductors are also shown.

It is interesting to note that, if we plot inductance of single (isolated) striplines as a function of their design width $w$, as shown in Fig. 13 by black symbols, the measured values fall perfectly on the simulated dependence, which uses the actual dielectric and metal thicknesses, the design linewidth, and the standard value of $\lambda = 90$ nm. This suggests that the real superconducting width of narrow Nb stiplines is smaller than their physical width measured in SEM by some amount $w_0$, which can be estimated as about 35 – 36 nm from Fig.13. This difference corresponds to a shift between the black solid curve and the red dash curve in Fig. 13.

The presence of a thin nonsuperconducting layer on the sidewalls of Nb lines can be a result of the surface oxidation or contamination by hydrogen known to suppress superconductivity of Nb. The thickness of this nonsuperconducting or weakly superconducting sheath can be estimated as $w_0/2 \sim 18$ nm on each sidewall of narrow Nb lines. For $w = 120$ nm isolated striplines on wafer 1, which measured $w_w = 114$ nm in SEM, the superconducting width, $w_s = w_w - w_0$, in this model would be only 78 nm. The simulated linear inductance of a 78-nm M6aM4bM stripline is 1.066 pH/μm, and 1.061 pH/μm if $w_s = 80$ nm. The measured inductance on this chip is $1.049 \pm 0.049$ pH/μm, see Fig. 9, agreeing with this model with <1.6% difference.

It is known that both oxidation and hydrogen absorption lead to Nb swelling due to increase in the unit cell volume. E.g., oxidation converts 1 nm of Nb into about 2.5 nm of $Nb_2O_5$ oxide. So, formation of an 18-nm oxide layer on the sidewalls, e.g., during wafer treatment in oxygen plasma, would require oxidation of only 7 nm of Nb. Since Nb lines are clamped to the substrate, their width at the interface with the underlying $SiO_2$ cannot change during the oxidation or hydrogen absorption and but the linewidth can increase on the free surfaces. As a result, the narrow lines should acquire a noticeable negative profile, having a larger linewidth at the top of the lines than at the bottom. It is also possible that thickness of the non-superconducting sheath $w_0/2$ depends on the density of lines. E.g., if hydrogen contamination happens during high density plasma etching as a result of sidewall bombardment by hydrogen or hydrogen-containing molecules and ions, dense photoresist pattern could release more hydrogen-containing molecules and produce more intense bombardment due to a higher negative charge accumulating on the photoresist/$SiO_2$ etch mask. The same would be true for oxidation in high-density plasma.

Many more experiments would be required to distinguish between the two possibilities described above: a relatively uniform bulk contamination of Nb (or increase in electron scattering) leading to a gradual increase in $\lambda$ with decreasing $w$ or a localized surface oxidation/contamination making the superconducting linewidth $w_s$ smaller than the physical width $w_w$.

### C. *Effect of Ground Plane Perforations on Inductance*

Adding perforations in the ground plane(s) below or/and above the signal conductor of microstrips and striplines is a well-known method of adjusting their inductance, capacitance, and impedance in microwave engineering; see, e.g., [47]-[49] and numerous references therein. It is also used in superconductor circuits for inductance adjustments; see e.g., [13], [23], [6]. The prior work considered very wide inductors and ground plane perforations which cannot be applied to modern circuits requiring much smaller dimensions and larger integration. In this work we consider inductors with a much smaller linewidth $w$ = 0.25 µm and smaller perforations as a practical example applicable to VLSI integrated circuits.

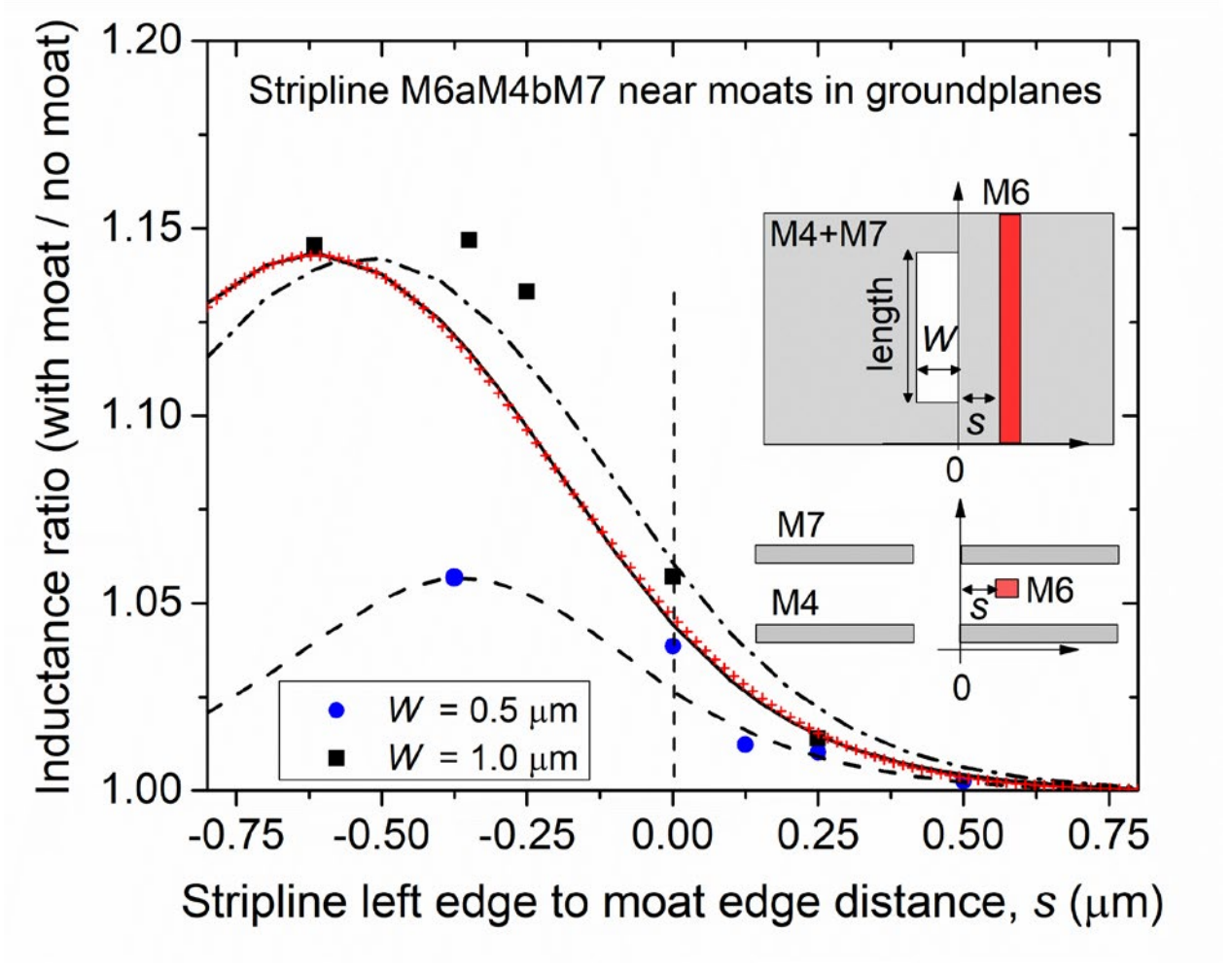

Fig. 15. Relative change in inductance of isolated stripline M6aM4bM7 as a function of distance, $s$ between the left edge of the signal conductor on layer M6 and the right edge of the congruent moats in M4 and M7 ground planes: dots (●) for the moats width $W$ = 0.5 µm; squares (■) for $W$ = 1.0 µm. The design linewidths of inductors on the layer M6 is $w$ = 0.25 µm. The measured inductance was normalized to the inductance value at $W$ = 0, corresponding to the unperforated ground planes. The simulated dependences using wxLL are also shown: dash line is for $W$ = 0.5 µm; solid line for $W$ = 1.0 µm; dash-dot line is for $W$ = 1.0 µm and a misaligned M7 moat shifted with respect to the moat in M4 by +0.1 µm towards the M6 conductor. In the latter case $s$ is the distance from the right edge of the moat in M4. A Gaussian fit (4) is shown by (+) symbols for $W$ = 1.0 µm. A cross section and top view of the inductors are shown for the congruent moats.

Fig. 14 shows the effect of congruent moats, long slits with width $W$ in the top and bottom ground planes, on linear inductance of a symmetrically placed stripline M6aM4bM7 and the results of simulations using wxLL software [13]. The data and simulations for M6aM4 microstips are also shown for a comparison. The data are normalized to the value of inductance at $W$ = 0, i.e., for unperforated ground plane(s). The width of the signal conductor on M6 layer is 0.25 µm.

In a practical range of the moats widths $W \le 2$ µm, inductance can be increased up to ~ 35% with respect to unperforated ground planes. Much wider perforations are impractical for dense circuits because the growth of the inductance slows down, inductance doubles only at $W \sim 10$ µm, whereas the circuit density suffers dramatically. The relative increase in inductance of the microstrips with $W$ is significantly smaller due to a larger distance to the nearest ground plane and a significant contribution of the space above the signal conductor to magnetic inductance.

Sometimes narrow bridges across long moats need to be placed for signal routing, transforming the long slit into a series of rectangular perforations. These bridged moats have a slightly smaller effect on inductance. For instance, for $w$ = 0.25 µm stiplines and $W$ = 2 µm moats with length 30 µm, the average increase in inductance $L/L(0)$ = 1.33 (Fig. 14), where $L(0) = L(\infty)$ is the inductance without the moat or at infinite distance from it. If instead of this long moat we use ten square perforations 2 µm x 2 µm spaced at 1 µm, the average increase in inductance of the striplines is 1.28. At a given moat width, the relative change in inductance $L/L(0)$ increases with increasing width $w$ of the signal conductor because narrower inductors have larger fringing fields which are less altered by the moat. For instance, in the previous case of ten square perforation, $L/L(0)$ = 1.33 if $w$ = 0.35 µm.

Ground plane perforations alter the return path of the superconducting current, forcing it to flow around the perforation instead of directly under the signal conductor. This can be modeled by inductance matrix containing self-inductances of the perforations and of the current paths along the signal conductor, returning along the perforations in the bottom and top ground planes, and mutual inductances between them [14]. Fortunately, inductance simulations using wxLL or wxLC, shown in Fig. 14, agree very well with the measurements, which simplifies the design procedure.

In the presence of the moats, inductance can be presented as

$$L(W) = L(0) + M(W), \tag{4}$$

where $M(W)$ is the aiding mutual inductance dependent on the moat width. Hence, the data in Fig. 14 correspond to $M(W)/L(0) + 1$. For the most practical moat width $W$ = 1 µm, the mutual inductance between the M6 stripline and congruent moats in M4 and M7 ground planes is $M \approx 0.15L(0) \approx$ 0.0855 pH/µm.

In superconductor digital circuits, congruent moats in the ground planes are widely used to collect magnetic flux from active areas of circuits and prevent trapping of Abrikosov vortices in superconducting films where they easily spoil performance of logic cells; see, e.g., [50] and references therein. Although absolutely necessary, these flux tapping moats unavoidably reduce circuit density and may also affect inductances of the nearby inductors due to mutual coupling of inductor to moats. To avoid this coupling, circuit designers place inductors quite far from the moats that decreases circuit density even further.

To study the effect of moats on inductors, we measured the relative change of stripline inductance as a function of distance, $s$ from the edge of congruent moats in M4 and M7 ground planes to the inductor edge, using dimensions typical for VLSI circuits. A sketch of the inductor cross section, showing mutual positions of the moats and the inductor is shown in Fig. 15. The results are shown in Fig. 15 as well. Negative values of $s$ correspond to the inductor moving into the moat; $s = -(W + w)/2$, where $W$ is the width of the moats, corresponds to a symmetrical position of the stripline along the middle of the moat, as in Fig. 14. We can see from Fig. 15 that the effect of moats on the inductance of nearby inductors is quite small − even for inductors with zero distance to the moats, $s$ = 0, the increase in the stripline inductance with respect to the stripline without moats is less than 6% for $W$=1.0 µm, and smaller for narrower moats.

The simulated linear inductance $L(s)$ has a symmetric bell-shaped dependence with the apex at $s_0 = -(W+w)/2$ where the center of the signal conductor is located symmetrically over the moats. Within and near the moat, this bell shape can be approximated, e.g., by a Gaussian as shown in Fig. 15 by (+) symbols,

$$L(s) = \Delta L \cdot exp[-(s+s_0)^2/(2d_{\text{eff}}^2)] + L(\infty), \quad (4)$$

where $\Delta L = L(s_0) - L(\infty)$ is the maximal moat-induced change of the inductance. The effective length $d_{\text{eff}}$, as well as $\Delta L$, depends on the moat width: $d_{\text{eff}} = 0.41$ µm and 0.31 µm, respectively, for $W = 1.0$ µm and $W = 0.5$ µm moats.

The data in Fig. 15 indicate that the moat influence on inductors diminishes very quickly, exponentially, with moving inductor away from the moat. This is easy to understand. The return currents in the ground planes flow mainly right under and above the signal conductor, spreading very little and decaying nearly exponentially with distance from the conductor edge with a typical decay distance of about $(d_1d_2)^{1/2} \sim 0.36$ µm, where $d_1$ and $d_2$ are the bottom and top dielectric thicknesses in the asymmetrical stripline. So, a moat in the ground planes placed farther away than this distance cannot significantly change the distribution of the return currents and, thus, should have negligible effect on stripline inductance and negligible moat to inductor coupling. Our measurements and simulation results demonstrate that spacing inductors from the moats by $\geq 0.25$ µm is sufficient to keep the moat effect on inductance $M(W)/L(0)$ below 3% for all practical moat widths $W \leq 2$ µm in the SFQ5ee and SC1/SC2 processes. Within the accuracy of the measurements, we were not able to detect any effect of the moats on stripline inductance at $s > 0.75$ µm. The simulation and the measurement accuracy was also not sufficient to distinguish between a simple exponential decay and (4) at $s > 0$.

The data points in Fig. 15 show some scattering, especially for inductors designed with $s = 0$, i.e., with coincident edges of the inductor and the moats. We think that this small variation is likely caused by misalignment of the fabricated inductors and the moat edges in the M4 and M7 ground planes. Misalignment between M6 and M7 has the largest effect because M7 is closer to the M6 inductor than M4. Assuming that M6 and M4 are perfectly aligned and that M7 moat is shifted by +0.1 µm, the simulated inductance is shown in Fig. 15 by a dash-dot bell curve for $W = 1$ µm moats. It demonstrates a few percent increase in inductance with respect to the perfectly aligned layers. Therefore, misalignment between circuit layers may be an additional source of small inductance variations for inductors in a close proximity to moats in ground planes.

To conclude this section, our inductance data indicate that coupling of inductors to distant, $s$ larger than about 0.75 µm, moats is extremely small. For $W = 1.0$ µm and $s = 0.25$ µm, mutual inductance with $w = 0.25$ µm stripline M6aM4bM7 was measured to be $M \approx 0.02L$ or about 0.01 pH/µm. If this moat traps one flux quantum $\Phi_0$, magnetic flux induced in stripline inductors at $s = 0.25$ µm should be only $0.02\Phi_0$ and much smaller for larger distances to the moats. This coupling should have negligible effect on the critical current of Josephson junctions in logic cells and the cells performance.

## V. Inductance of Vias Between Layers

### A. Via Structure and Measurements

Superconducting vias are an essential component of all superconductor integrated circuits, providing connections between components located on different layers and closing superconducting loops forming circuit inductors. Reduction of the size of vias is a much more difficult task than reduction of the linewidth of inductors and is a subject of considerable technological efforts needed to increase the integration scale of SCE [51], [52]. Contribution of vias, especially of submicron sizes, to the inductance of superconducting loops formed in integrated circuits is largely unknown. Below we present experimental data for many configurations of vias in the SFQ5ee [30] and SC1/SC2 processes [5].

Vias I5 connecting layers M5 (JJ base electrode) and M6 are used very frequently because they are required to form resistively shunted junctions (RSJ) in the SFQ5ee and SC1/SC2 processes and inductors on M6 layer. Any via presents typically a rectangular opening in the interlayer dielectric, I5 in this case, filled by niobium of the next layer, M6 in this case. The slope of the sidewall is about 75 degrees. Bottom dimensions of the opening are $a$ and $b$ in the perpendicular and parallel to the current directions, respectively; square vias are used most frequently. The typical SEM image of I5 via cross section with $a = b = 700$ nm is shown in Fig. 16. The opening in the dielectric is surrounded by Nb of the bottom and top layers by some amount, $sr$, in order to increase via reliability if there is a misalignment between the layers. The combined object which we refer to as the via has dimensions $a+2sr = w$ perpendicular to the current direction and $b+2sr$ along the current direction.

To increase the accuracy of the measurements, we used chains of vias containing from 20 to 64 vias and placed them in a stripline configuration, e.g., with M4 and M7 ground planes most often used in circuits, as shown in Fig. 16. Parasitic inductance of the ground connection at the chain end, using three staggered vias I6,I5, and I4 shown in Fig. 16, was deducted using the differential scheme described in II.B.

A chain containing an even number $n$ of I5 vias has $n/2 - 1$ pieces of M6 wires and $n/2$ pieces of M5 wires each of length $s$, connecting vias in the chain. To find the inductance per single via, $L_{\text{via}}$ we deducted the contribution of connecting wires from the total measured inductance, using two methods: a) by deducting inductances of the stripline M6aM4bM7 with length $s\cdot(n/2 - 1)$ and width $w = a + 2sr$, and of the stripline M5aM4bM7 with length $s\cdot n/2$ and the same width $w$, calculated using the stripline inductance data measured on the same chip and given in Table III; b) by including these two striplines and two I5 vias as the left inductor in the differential measurements scheme in Fig. 2. By plotting the resultant bare inductance versus the number of vias and using the linear fit $L_{\text{chain}} = n\cdot L_{\text{via}}+L_0$, as shown in Fig. 17, we determined the inductance per via, $L_{\text{via}}$. In all cases, we observed that the residual term $L_0$ was negligibly small in comparison with the measured inductance, indicating that the differential method of removing parasitic contributions worked very well. A similar procedure was used for all other types of vias.

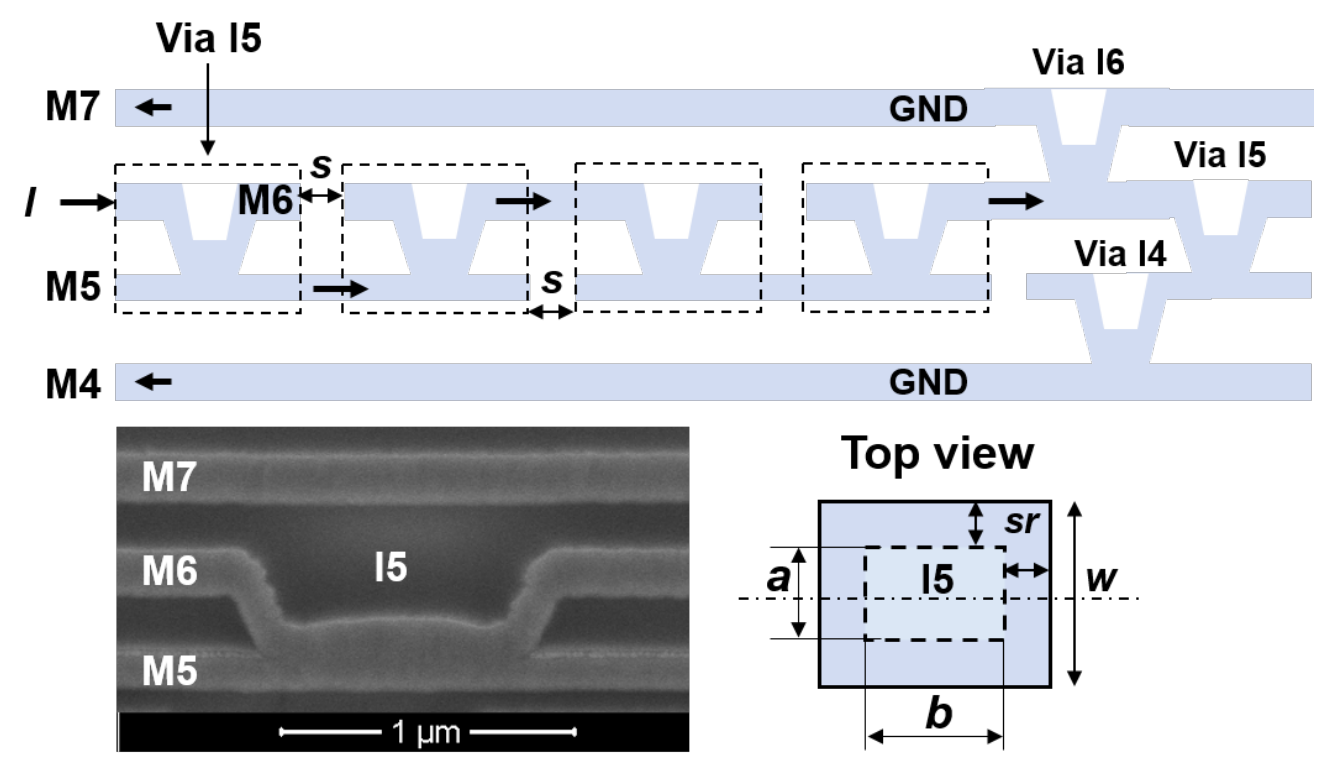

Fig. 16. Schematic cross section of a chain of four I5 vias between M4 and M7 ground planes, showing direction of electric current alternating between M6 and M5 layers. The current is returned to the ground planes through staggered vias I6, I5, and I4 connecting M7 and M4 layers. A via presents an opening in the interlayer dielectric, with dimensions *a* and *b* at the bottom, filled by the next Nb layer as show in the SEM image of the cross section. The opening must be surrounded by the top and bottom Nb by *sr* on all sides for via reliability in case of misalignment. Top view shows dimensions *a*, *b sr*, and *w*. In the chain, adjacent vias are spaced by *s* and interconnected by short pieces of wires with width *w* in alternating layers to form a meandering current path. Much longer chains containing from 20 to 60 vias were used in inductance measurements to increase the accuracy. The contribution of the ground connection on the right, by vias I6, I5, and I4, was deducted using the differential method. The contribution of metal wires connecting the vias in M5 and M6 layers was also deducted to find inductance per single via; see text.

### *B. Inductance of Vias Between Superconducting Layers*

The obtained results are summarized in Table III. We present a subset of data for two wafers, w6 and w8, fabricated in the same run of the SC1 process and two wafers, w3 and w4, fabricated in a different run in order to characterize the cumulative repeatability of the fabrication process and the measurements. The data should be also valid for the SFQ5ee process due to a close similarly in the layer thicknesses, and we included in Table III some data obtained on wafers made in SFQ5ee process for comparison.

Via inductance crucially depends on the current return path, i.e., the type and number of ground planes, because inductance is electromagnetic property of a loop. So, the data in Table III should only be used in this context. For instance, if two inductors M6aM4bM7 and M5aM4bM7 are connected by an I5 via it is appropriate to add $L_{\text{via}}$ from Table III to the total loop inductance. In asymmetrical inductors like these, a larger return current flows in the closest to the signal conductor ground plane than in the more distant one, and magnetic field is higher between the signal conductor and the closest ground plane. The closest ground plane is different for different inductors, e.g., it is M7 for M6 inductors and M4 for M5 inductors. When M6 and M5 inductors are connected, the return currents and magnetic field need to redistribute in the region of the via, contributing to the total inductance of the via.

Although vias are three-dimensional objects, we can gain some insight into their inductive properties from the properties of individual films and the data in Table III. We see that, at a constant size *a* and wire width *w*, via inductance linearly depends on its length in the direction of applied current, *b*, $L_{\text{via}}=L_0+L_1\cdot b$; see inset in Fig. 17. By inspecting the via cross section in Fig. 16, the current path in the via, from left to right, consists of a flat piece of M6 wire with length *sr* in front of the opening in the dielectric, followed by a front wall of the via where most of the current flows down, similar to a waterfall, and a piece of wire with length *b* and width *a* at the bottom of the via, and then M5 wire with width *w* and length *sr*. Inductance of the via can be estimated as the sum of these three components:

$$L_{\text{via}} = (\ell_{M5} + \ell_{M6})\cdot(w-a)/2 + L_{\text{wall}} + b\cdot\ell_{M5M6}, \qquad (5)$$

TABLE III
INDUCTANCE OF VIAS OF VARIOUS TYPES AND DIMENSIONS

| Via or Wire | Grnd planes | *a* (nm) | *b* (nm) | *sr* (nm) | $L_{\text{via}}$, w6 (pH) | $L_{\text{via}}$, w8 (pH) |
|---|---|---|---|---|---|---|
| I5 | M4+M7 | 700 | 700 | 350 | 0.245 | 0.245 |
| I5 | M4+M7 | 700 | 700 | 150 | 0.287 | 0.281 |
| I5 | M4+M7 | 500 | 500 | 250 | 0.318 | 0.325 |
| I5 | M4+M7 | 400 | 400 | 300 | 0.362 | 0.362 |
| I5[a] | M4+M7 | 500 | 500 | 250 | 0.303 | 0.317 |
| I5[a] | M4+M7 | 500 | 1000 | 250 | 0.389 | 0.394 |
| C5[a] | M4+M7 | 500[b] | 500[b] | 250 | 0.314 | 0.307 |
| I6 | M4+M9 | 500 | 500 | 250 | 0.394 | |
| I6 | M4+M9 | 400 | 400 | 200 | 0.455 | |
| I6 | M4+M9 | 400[c] | 400[c] | 200 | 0.473[b] | |
| I6 | M4+M9 | 350[d] | 350[d] | 175 | 0.484 | |
| I7 | M4+M9 | 500 | 500 | 250 | 0.393 | |
| I6 | M4 | 500 | 500 | 250 | | 0.705 |
| I6 | M4 | 400 | 400 | 200 | | 0.876 |
| I6 | M4 | 400[b] | 400[b] | 200 | | 0.918[b] |
| I6 | M4 | 350[c] | 350 | 175 | | 0.927 |
| I5I6I7 | M4+M9 | 500 | 500 | 250 | 1.249[e] | |
| M5 | M4+M7 | | | | 0.3094[f] | 0.3006[f] |
| M6 | M4+M7 | | | | 0.2596[f] | 0.2569[f] |
| M6 | M4 | | | | | 0.6405[f] |
| M6 | M4+M9 | | | | 0.3846[f] | |
| M7 | M4 | | | | | 0.7338[f] |
| M7 | M4+M9 | | | | 0.3969[f] | |
| M8 | M4+M9 | | | | 0.3163[f] | |
| | | | | | w3[g] | w4[g] |
| I7 | M4+M9 | 500 | 500 | 250 | 0.365 | 0.342 |
| I7 | M4+M9 | 500 | 750 | 250 | 0.412 | 0.388 |
| I7 | M4+M9 | 500 | 1000 | 250 | 0.472 | 0.443 |
| I7 | M4+M9 | 1000 | 500 | 250 | 0.274 | 0.256 |
| I6I7[h] | M4+M9 | 500 | 500 | 250 | 0.651 | 0.596 |
| I6I7[i] | M4+M9 | 500 | 500 | 250 | 0.747 | 0.686 |

[a]Wafer fabricated in the SFQ5ee process [30]
[b]Diameter of a circular via to Josephson junction on layer J5
[c]Circular via diameter; width of connecting wires $w = a+2sr = 800$ nm
[d]Width of connecting wires $w = a+2sr = 700$ nm
[e]Via comprised staggered vias I5, I6, and I7, connecting layers M5 and M8
[f]Inductance per unit length in pH/μm of wires with width $w = 1$ μm and the specified ground planes, measured on the same chip and used in the calculations of via inductance; see text
[g]Wafers w3 and w4 are from a different fabrication run of the SC1 process [5]
[h]I6 and I7 vias are spaced at zero distance in the direction of current
[i]I6 and I7 vias are spaced at 250 nm in the direction of current

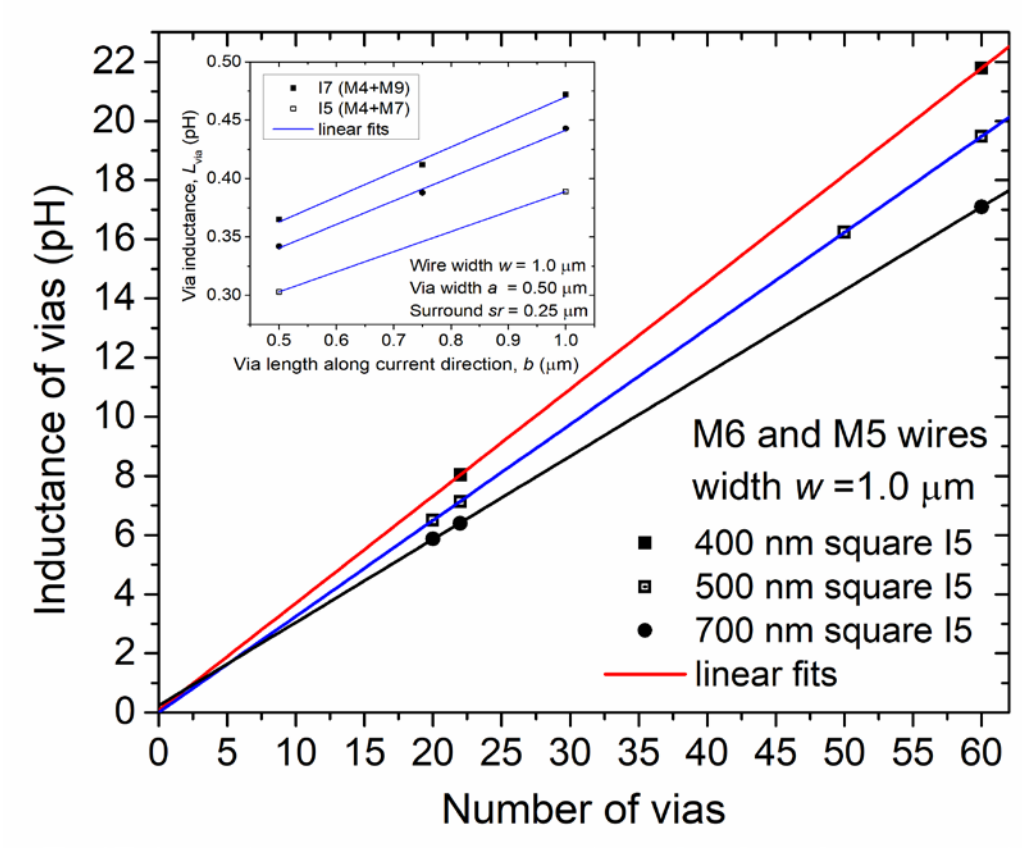

Fig. 17. Inductance of chains of square vias I5 with different number of vias in the chain, and for different via sizes, $a$. Inductance of interconnecting wires M5 and M6 was subtracted from the total inductance of the chain using differential measurements or independently measured inductance per unit length of M5 and M6 wires with $w = 1$ µm between M4 and M7 ground planes. The slope of the linear fit gives inductance per via, $L_{via}$ shown in Table III. Inset shows inductance per via as a function of the via length $b$ along the current direction for I7(M4+M9) and I5(M4+M7) vias with $a$ = 500 nm. Solid lines are linear fits to $L_{via}=L_0+L_1\cdot b$, giving the averaged values $L_0 = 0.248\pm0.01$ pH and $0.229\pm0.01$ pH, respectively, for I7(M4+M9) and I5(M4+M7) vias, and $L_1 = 0.208\pm0.01$ pH/µm and $0.163\pm0.01$ pH/µm, respectively, for the same types of vias.

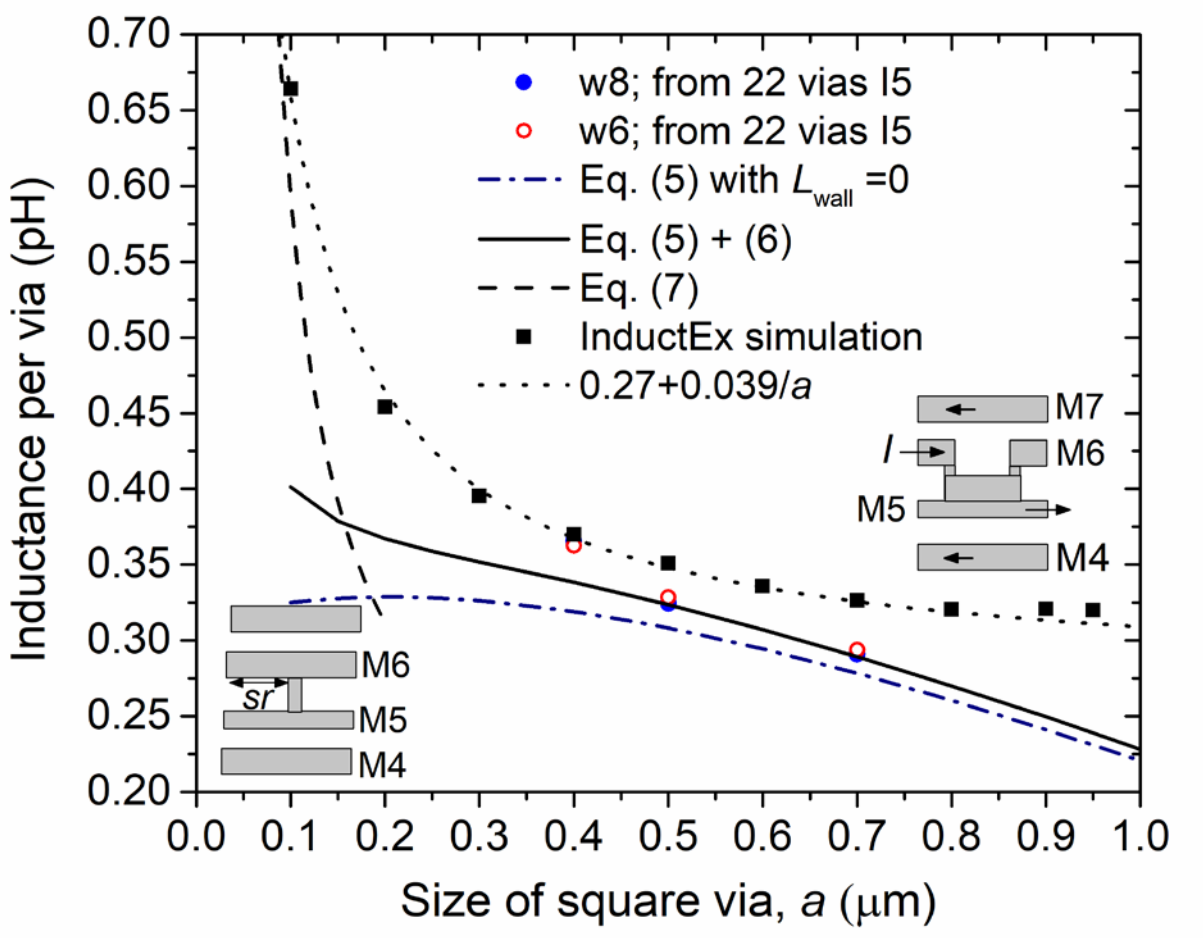

Fig. 18. Simulated inductance of I5 square vias connecting layers M6 and M5, above M4 and below M7 ground planes, with the model cross section shown on the right: dash-dot curve is (5) at $L_{wall} = 0$; solid curve is (5) with $L_{wall}$ given by (6); dash curve is an asymptotic inductance (7) for pillar vias made by a stud-via or damascene process with small, $a < 2\lambda$, diameters as sketched in the left bottom corner. Via inductance was measured using test structures containing 22 square vias I5 and shown for two wafers in Table III, wafer 6 (○) and wafer 8 (●) for several via sizes, $a$ at a fixed width of M5 and M6 wires, $w = 1$ µm. Simulations using InductEx for the same structures are shown by (■); dotted curve is a simple approximation of the InductEx simulations by a hyperbolic dependence, and mainly shown to guide the eye. A smooth transition between the solid and dash curves is expected from a full model.

where we used that $sr = (w-a)/2$. In (5), parameters $\ell_{M5}$ and $\ell_{M6}$ are linear inductances of, respectively, M5 and M6 wires with width $w$, given in Table III; $\ell_{M5M6}$ is linear inductance of the via bottom layer, and $L_{wall}$ is inductance associated with the current dropping to the bottom of the via. There can be other terms related, e.g., to current crowding near the front wall and at the exit from thick bottom into the thinner M5, but we will neglect them for simplicity.

The metal thickness at the bottom of large vias is $t_{M5M6} = t_{M5}+t_{M6}$, the sum of M5 and M6 thicknesses given in Table I. It is slightly less in small vias due to shading effects during the top metal deposition into the opening. We will neglect this difference. Then, we simulated the linear inductance $\ell_{M5M6}$ for different linewidths $a$, using wxLL. Using the simple model (5), we calculated dependence of via inductance on via size $a$ for square $a$=$b$ vias, shown in Fig. 18 by a dash-dot line for $L_{wall} = 0$, along with the measured data from Table III.

From via cross section, we determined that Nb thickness on via sidewall, $t_w$ is about 40% of the layer thickness on flat surfaces, about 80 nm for I5 vias, which is less than λ. The height of the drop is about the difference between the interlayer dielectric thickness and the top metal layer thickness, $d_{M5M6} - t_{M6} \approx 60$ nm for I5 vias. Then, the contribution of the front wall drop to the inductance can be estimated assuming that this inductance is mainly kinetic, giving

$$L_{wall} = \mu_0(\lambda^2/t_w)\cdot(d_{M5M6} - t_{M6})/a, \qquad (6)$$

where $\mu_0\lambda^2/t_w$ is kinetic inductance per square and $(d_{M5M6} - t_{M6})/a$ is the number of squares of Nb film on the front wall of the via. Including (6) into (5) results in the dependence of via inductance on via size shown by solid curve in Fig. 18, which nicely agrees with the data, especially at sizes $a \geq 0.5$ µm.

The via cross section in Fig. 16 and the described simple model cannot be applied to vias smaller than about 400 nm for two reasons. Firstly, openings in the dielectric with sizes comparable to the dielectric thickness cannot be filled by metal sputtering because of clogging near top of the opening. Secondly, the opening shape becomes very rounded, rather circular than square. For making high aspect ratio vias, completely different approaches needs to be used, for instance, a stud-via process described in [50] where cylindrical metal pillars (studs) are etched and then interlayer dielectric is deposited over them and planarized by CMP. Another one is a damascene process filling in the openings by a metal chemical vapor deposition to form cylindrical pillars and then removing the metal deposited on the flat surface by a metal CMP.

In any of these advanced processes, via would present a superconducting rod with diameter $a$ connecting two superconducting wires on different layers. For deep submicron via diameters, which are the main focus of these advanced processes, the rod inductance can be estimated using a formula similar to (4) where kinetic inductance of the rod $L_k = (4\lambda^2/\pi a^2)d$ will dominate; $d$ is the rod height equal to the dielectric thickness. In this asymptotic case, the via inductance becomes

$$L_{via} = (\ell_{M5} + \ell_{M6})\cdot(w-a)/2 + (4\mu_0\lambda^2/\pi a^2)d \qquad (7)$$

This dependence is shown in Fig. 18 by a dash curve. A full dependence of $L_{via}$ on via size should provide a smooth transition from the dash-dot curve (5)-(6) for relatively large vias with $a > 0.4$ µm to the dotted curve (7) for $a < 2\lambda$. To summarize this discussion, we developed a simple description of via inductance based on inductive properties of superconducting

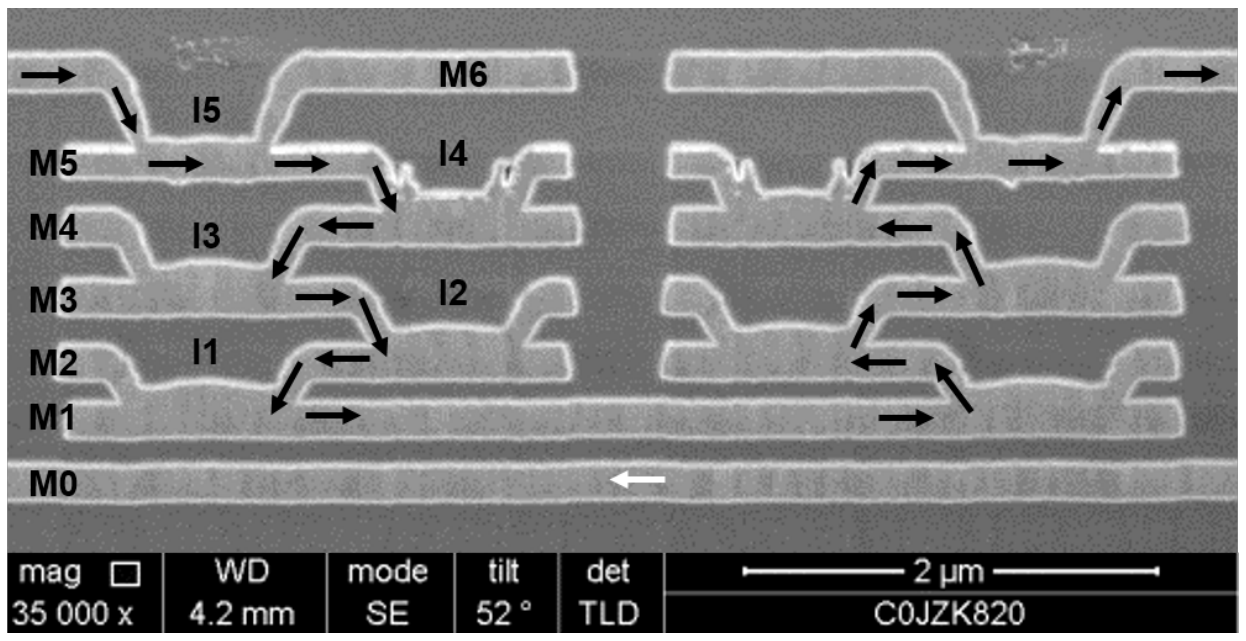

Fig. 19. Cross section of a chain of composite stacked-staggered vias I1I2I3I4I5 connecting layers M1 and M6, over M0 ground plane, fabricated in the SFQ5ee process. Vias $I_i$ connecting neighboring layers $M_i$ and $M_{i+1}$ are offset by $2\cdot sr$ from vias $I_{i+1}$ on the next level connecting $M_{i+1}$ and $M_{i+2}$. Individual vias are square with $a = 700$ nm, with surround $sr = 350$ nm, and spacing between adjacent vias $I_i$ and $I_{i+1}$ and between the stacks of 500 nm.

TABLE IV
INDUCTANCE OF STACKED-STAGGERED VIAS IN SFQ5EE PROCESS

| Via | Grnd planes | *a, b* (nm) | *sr* [a] (nm) | $L_{via}$, w3[b] (pH) | $L_{via}$, w7[b] (pH) |
|---|---|---|---|---|---|
| I1I2I3I4I5 | M0 | 700 | 350 | 1.423 | 1.399 |
| I1I2I3I4 | M0 | 700 | 350 | 1.107 | 1.109 |
| I5 | M0 | 700 | 350 | 0.316 | 0.290 |
| I1I2I3I4I5 | M0 | 500 | 350 | 1.587 | 1.519 |
| I1I2I3I4/*n* | M0 | 700 | 350 | 0.277 | 0.277 |
| I1I2I3I4I5/*n* | M0 | 700 | 350 | 0.285 | 0.280 |
| I1I2I3I4I5/*n* | M0 | 500 | 350 | 0.317 | 0.304 |

[a]Spacing between the individual vias was 500 nm instead of $2\cdot sr$ =700 nm in order to minimize the area of composite vias

[b]Wafers w3 and w7 are from two different fabrication runs of the SFQ5ee process [30]

films comprising vias and obtained a good agreement with the available data.

Because vias are truly three dimensional objects, we also attempted to simulate inductance of I5 via chains using 3D inductance extractor InductEx [19]. We used a layer definition file corresponding to the fabrication process and circuit layouts corresponding to the differential measurements described above. These layouts have $n = 24$ via chains in the right arm of the SQUID in Fig. 2. In its left arm the SQUID has an I5 via to a 6-µm-long M5 wire, length $sr\cdot n/2$, followed by another I5 via connecting a 5.5-µm-long M6 wire, length $sr\cdot(n-1)/2$, thus allowing to automatically deduct contributions of via interconnecting wires in the right arm and all ground connections. The differential number of vias in the layouts is 22. The version of InductEx used in simulations assumes a vertical slope of via sidewalls.

For the linear inductance of 1-µm-wide M5aM4bM7 and M6aM4bM7 striplines, InductEx gives 0.3041 pH/µm and 0.2576 pH/µm, respectively, which closely, within 0.3%, matches the average of values for the two wafers of 0.305 pH/µm and 0.258 pH/µm in Table III, passing the first check. The simulated inductance per I5 via is shown in Fig. 18 by black squares for various sizes of square I5 vias at the fixed width of wires $w = 1$ µm. The agreement with the measurements is not good as it appears that the experimental dependence and the simulated one are different and simply intersect at one point at $a\sim 0.4$ µm. Surprisingly, InductEx cannot find a solution if $a = w$, a completely legitimate case from mathematical standpoint and often encountered in semiconductor electronics manufacturing. In this case all the current flows down the front sidewall of the first via and up the front sidewall of the second via, and so on, in chains of vias. The asymptotic behavior at $a \to w$ indicates a value of 0.32 pH per via, which is significantly larger than the inductance of a 1-µm long piece of wire with thickness $t_{M5}+t_{M6}$, above M4 and below M7 ground planes. The dotted black curve in Fig. 18 is our approximation of the InductEx simulated values by a simple dependence in order to guide the eye. More work is required to understand the source of these differences, improve the software, and develop models for small vias.

## C. *Inductance of Staggered and Stacked-Staggered Vias*

Composite vias comprising several individual vias placed at the minimum allowed spacing are used to connect distant layers, e.g., via I4I5I6 connecting M4 and M7 ground planes is sketched in Fig. 16. Vias between adjacent pairs of metal layers, $I_i$ and $I_{i+1}$, are offset (staggered) to make contact on the planar surface of metals $M_i$ and $M_{i+1}$. The next level via $I_{i+2}$ can be either also offset in a staircase fashion or stacked on top of $I_i$ in order to minimize the total area. A cross section of a stacked-staggered via I1I2I3I4I5 connecting layers M1 and M6 is shown in Fig. 19 as an example. Usually, the offset between vias is equal $2\cdot sr$, allowing to assemble composite vias from individual parametric via cells.

In double vias, like I5I6 or I6I7, the transport current always flows in the same direction in both vias. Therefore, our simple model presented in III.B should apply, and inductance of the double via should be equal to the sum of inductances of the individual vias. The data in Table III more or less support this. Similarly, for a staircase arrangement of $n$ vias, e.g., I1I2I3…, inductance of the composite via should be approximately equal to the sum of inductances of individual vias. In stacked-staggered vias, electric current in adjacent layers flows in the opposite directions in a meandering fashion, as shown by arrows in Fig. 19. This creates a negative mutual inductance between the layers and forces the current to flow near only one wall of the vias without spreading out along the bottom of the vias. In this case, we could expect the total inductance to be a bit lower than the sum of individual inductances.

We measured inductance of the stacked-staggered vias shown in Fig. 19 in order to provide data for circuit design and model building. Test circuits containing chains of stacked-staggered vias were fabricated in SFQ5ee process [30], and we used the approach described in V.B to infer inductance per composite via. The data for stacked-staggered vias comprised $n$=4, for I1I2I3I4 via, and $n$=5, for I1I2I3I4I, individual square vias are shown in Table IV. First of all, we note that inductance of these composite vias is quite substantial. A typical inductor in logic cells of superconductor integrated circuits has a loop inductance on the order of 10 pH, and two vias are required to form a loop between different layers. So the inductance of two composite vias can reach up to 20% - 30% of the typical loop inductance and, hence, needs to be accounted for quite accurately. Similarly, if M1 layer is used in a PTL to transmit data between logic cells, inductance of the M1 to M6 path, i.e., of via I1I2I3I4I5, needs to be taken into account.

Individual vias are connected in series in the composite via. So, we can estimate a contribution of the top I5 via by subtracting inductance of I1I2I3I4 via from inductance of I1I2I3I4I5 via. The result is given in the third row of Table IV; it does not differ significantly from the I5 via inductance in Table III, obtained using chains of individual I5 vias. We separated I5 vias because they slightly differ from all other vias in the stack because I5 dielectric thickness is 280 nm in the SFQ5ee process whereas thickness of all other dielectric layers is 200 nm; see Table I. We can simply characterize the average contribution of individual vias by dividing the total inductance into the number $n$ of individual vias, as shown in Table IV. The difference between $n$=4 and $n$=5 cases is within the error of the measurements and the wafer-to-wafer variation. Therefore, we recommend to use in circuit design the following average inductance values for 700-nm square vias in the SFQ5ee process: 0.28 pH per via for I0, I1, I2, I3, I4, and I6, and 0.29 pH per via for I5 vias. Similarly, for 500-nm square vias, we suggest values of 0.32 pH for I5 and 0.31 pH per via for all other vias. The difference between inductance of vias with $a$ = 0.5 µm and 0.7 µm is not that significant.

## VI. Inductance of Thin Film Resistors

Thin-film resistors are used to shunt Josephson junctions in order to damp Josephson oscillations and control JJ switching dynamics. Shunt resistors operate with ac currents of very high frequencies up to about 1 THz, contrary to bias resistors which are used to simply set dc bias current through JJs. Inductance of shunt resistors can strongly affect JJ dynamics because it can resonate with junction capacitance and cause many nonlinear effects [53]-[55]. Hence, knowledge of this inductance and ability to control it is important for integrated circuit design and fabrication process development.

Geometrical inductance of thin-film resistors can be simulated using inductance extraction software like FastHenry [20] and InductEx [19], and is usually relatively small due to the presence of superconducting layers, ground planes, below and above the resistors. In [56], a relatively large inductance associated with thin-film shunt resistors was found and ascribed to the imaginary part of the complex conductivity and the associated kinetic inductance of thin normal-metal films at frequencies larger than electron scattering frequency. This interpretation may require an additional re-examination which is beyond the scope of this paper and will be done separately.

Inductor extractors [19], [21] do not account for the kinetic inductance of resistors because they use only the real part of the complex conductivity of thin films at finite frequencies in their computational engines. In S. Whitely's [21] adaptation of the FastHenry [20] to superconductors, the complex conductivity is used to treat Meissner effect in superconductors; see, e.g., [57], [58]. Hence, extending this approach to thin-film resistors in the normal state should be straightforward.

## VII. Discussion

### A. Flux Trapping and Coupling to Moats

Recent flux trapping experiments, using M6aM4 microstip inductors and circuits fabricated in the SFQ5ee process at MIT LL, found a much stronger coupling between microstrips and moats than what we measured in this work; see [59] and references therein. This is partly expected because, due to the large distance $d_1$= 0.615 µm between M4 and M6 layers, see Table I, the return current spreads to much larger distances in these microstrips than in the striplines with two ground planes where $d_2$ = 0.2 µm mainly determines the current spreading around the moats. However, in [59] a coupling was observed to moats located at distances much farther away than would be consistent with the results of our measurements showing a very fast, nearly exponential decay of coupling with distance to long and narrow moats. We suggest that coupling to distant moats observed in [59] is an artifact of the experimental procedure used as well as a result of a difference between microstrip configuration used in [59] and stripline configuration used in this work.

To explain these differences, consider a simplistic circuit containing a ground plane with two moats and a microstrip as sketched in Fig. 20a. In a typical experiment with superconductor integrated circuits, a circuit is cooled down in a weak, but unfortunately unavoidable, residual magnetic field $B_r$, which source is far away from the chip. The source can be the Earth field or not completely demagnetized mu-metal shields [60], [61], and is represented in Fig. 20a by two poles of a magnet. Above the superconducting critical temperature, $T_c$ the field distribution in the ground plane represents the field source pattern and is typically more or less uniform. All the field lines start and end on the magnet poles, and the field does not change direction within the circuit.

The purpose of the moats is to attract all Abrikosov vortices which can form in the ground plane upon field cooling or make their formation energetically unfavorable by reducing the effective width of the ground plane strips between long moats below the critical width $w_c = (2\Phi_0/\pi B_r)^{1/2}$; see [50] and references therein. Cooling through the $T_c$ should be very slow in a narrow temperature range where the moat attraction force is larger than the vortex pinning by film defects, thus allowing vortices to diffuse to moats before a rapidly growing with decreasing temperature pinning force makes them immobile. Moats length $l$ should be large, $l >> \lambda(T)^2/t$, in order to provide a sufficient attraction force for vortices in that narrow temperature range near the $T_c$. Because very near $T_c$ the effective penetration depth $\lambda(T)^2/t$ determining the range of screening currents is quite macroscopic, small moats do not perturbe the current distribution near the vortex substantially, i.e., do not create a substantial gradient of the superfluid density, to generate a noticeable attraction force.

If these cooling conditions fulfilled, below the $T_c$, magnetic flux is completely expelled from the ground plane to outside of the chip and into the moats as shown schematically in Fig. 20a. Superconducting screening currents are induced around edges of the moats and around the periphery of the ground plane. These currents exponentially decay in films with thickness $t > 2\lambda$. Therefore, a microstrip placed a few λ away from the moat edges is screened from the flux trapped in the moats; magnetic flux induced by the current in the microstrip does not couple to the flux trapped in the moats. This situation basically corresponds to the experimental conditions used in this work, except that we used striplines.

Now consider what happens if the external magnetic field is switched off, as was a part of the experimental procedure in [59]. Magnetic flux outside of the chip vanishes, but it cannot disappear from the moats in the superconductor (or from the vortices trapped in the film), because this would require destruction of superconductivity of the ground plane. However, magnetic field lines must be closed according to the Maxwell equations. Therefore, magnetic field lines must rearrange to form closed loops, closing around the ground plane and through the moats, as shown schematically in Fig. 20b. This completely changes configuration of the magnetic field around the circuit and distribution of currents in the ground plane. Now, the field lines and the superconducting screening currents need to spread out over the ground plane and appear also on its bottom side. This spreading out increases coupling between microstrips on both sides of the ground plane to the nearest moats and induces coupling to distant moats. Apparently, this artificially induced coupling is what was observed in [59].

Yet more adverse situation takes place if a circuit field cooling is done in a nonuniform, sign-changing field of a local source, e.g., a wire with current $I_{mag}$, placed near the chip surface, as was used in [59] and depicted schematically in Fig. 20c. Above $T_c$, the closed-loop field lines produced by $I_{mag}$ penetrate the ground plane at various angles; the field decays inversely proportional to the distance from the wire and changes direction on its opposite sides. Below $T_c$, the flux lines become expelled into the moats in such a way that moats on opposite sides of the wire have opposite directions of magnetic flux and superconducting currents are induced around and between the moats linked by the common flux. After the field created by $I_{mag}$ is tuned off, magnetic flux remains trapped in the moats, Fig. 20d. The amount of magnetic flux in the moats decays slowly, approximately inversely proportional to the distance from the wire, and all moats in the ground plane, even very distant ones, become coupled by the flux lines closing though the moats. As a result, any inductor above and/or below the ground plane becomes coupled to all moats in the ground plane. This explains why and how inductor-to-moat coupling observed in [59] occurs. An artificially created flux distribution in and around the ground plan induces screening currents between the moats and between the moats and periphery of the ground plane, which couple to inductors placed even far away from the moats and the chip edges. This coupling does not occur in correctly done experiments with superconductor integrated circuits, in which residual magnetic field does not change directions within the circuit and is never changed after chip cooling below the $T_c$.

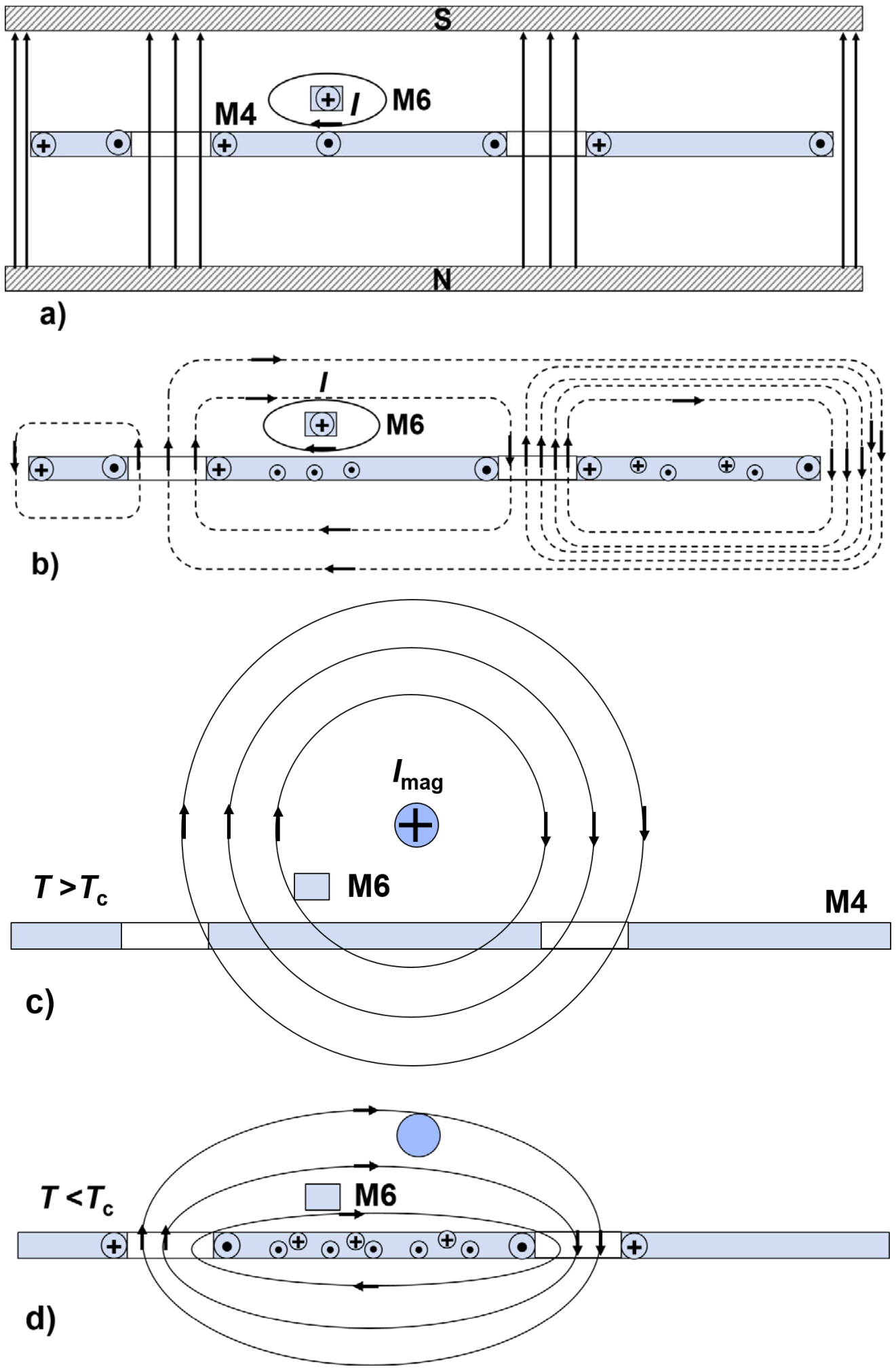

Fig. 20. Schematic cross section of a microstrip M6aM4, as an example, with two flux trapping moats and distribution of magnetic field lines corresponding to different experimental conditions.
(a) After cooling in a weak magnetic field created by a distant source. Magnetic flux is expelled from the superconducting ground plane and trapped in the moats. Superconducting Meissner currents are flowing around the moats with trapped flux and periphery of the ground plane. The currents are confined to distances on the order of magnetic field penetration depth $\lambda$. The field lines start and end on the magnet (residual field source) poles.
(b) After the field is switched off. The field lines must form closed loops, closing through moats and around the ground plane, as shown schematically. The trapped flux in the moats remains unchanged but the field near and current distributions in the ground plane need to change substantially, causing coupling between inductor M6 and the moats. Superconducting currents are now induced everywhere in the ground plane, both on the top and bottom surfaces of the ground plane. Magnetic flux in the moats becomes linked to the loop formed by the microstrip M6 and the ground plane.
(c) The same experiment now done using a local field sources, e.g., a wire with current $I_{mag}$ or a coil placed near the circuit. The field direction changes sign within the circuit. In the normal state, field lines are closed loops going through the ground plane and around it.
(d) Below $T_c$, magnetic flux is expelled into the moats, forming fluxons of the opposite polarity with screening currents circulating around the moats and between them on both sides of the ground plane. A much stronger coupling to the inductor M6 is induced in this case than in the a) - b). Switching the field off changes the field and current distribution around the periphery of the ground plane but does not change it near the inductor. In this case, coupling of nearby inductors to the moats is the result of an artificially created bi-polar distribution of the local magnetic field that changes directions within the circuit, on opposite sides of the field source.

### B. Moats and Flux Trapping: Striplines vs Microstrips

Another major difference between the measurements in this work and in [59] is that we used stripline inductors with two connected ground planes. This configuration provides a much better shielding than a single ground plane, even in the worst case considered above. Indeed, upon a slow cooling of the chip, the flux is expected to be trapped in the congruent moats as shown schematically in the cross sections in Fig. 21a. If the field is somehow switched off, the flux lines need to rearrange into closed loops. However, in this case they can only close outside of the top ground plane M7 due to the presence of superconducting vias connecting the top and bottom ground planes; see Fig. 21b. Now, the screening currents will appear on the bottom surface of the bottom ground plane M4 and on the top surface of the top ground plane M7. However, the possibility of the field lines slipping under the M7 and affecting the inductor M6 is very small.

If the field was initially parallel to the ground planes, the flux may remain frozen in superconducting loops formed between vias connecting M4 and M4, and may still affect cell inductors and junctions between M4 and M7.

This discussion summarizes why a single ground plane and microstip inductors are rarely used in modern superconductor integrated circuits. They are a feature of the past when the number of superconducting layers was strongly limited by rudimentary fabrication technologies. Stripline configurations with two and more ground planes provide much better circuit shielding and protection against flux trapping and parasitic coupling to moats.

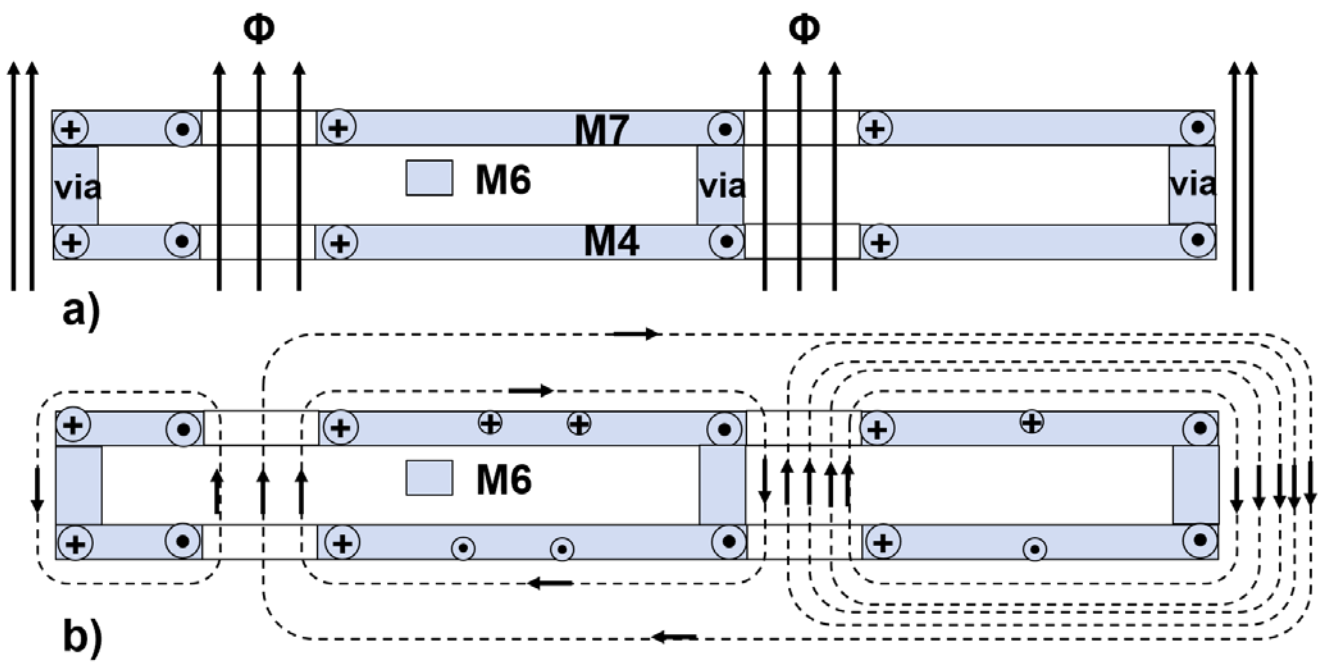

Fig. 21. Schematic cross section of a stripline M6aM4bM7 with flux trapping moats and distribution of magnetic field lines corresponding to the experimental conditions used in this work. The M4 and M7 ground planes are connected using superconducting vias placed on the periphery of the ground planes and between the moats, forming closed superconducting loops around the stripline. The effect of the nearest moat on the left was studied. This moat is not separated by vias connecting M4 and M7, whereas a more distant moat on the right is.
(a) After field cooling in a weak residual field of the mu-metal shields, magnetic flux is expelled and trapped in the moats. The field cannot be switched off. All bias currents to the test circuit were applied and removed while the circuit was in the superconducting state, which cannot change flux distribution in the moats of the circuit.
(b) If the field is switched off, a redistribution of flux and currents must happen to form closed B-field lines. However, in this case, the currents are mainly induced on the top surface of the top ground plane M7 and the bottom surface of the bottom ground plane M4. The stripline coupling to the flux in the moats remains weak and unchanged from (a). Coupling to the moat on the right is negligibly small due to vias.

## VIII. Conclusion

In conclusion, we presented inductance data for many features commonly encountered in modern superconductor integrated circuits such as straight line inductors in microstrip (one ground plane) and stripline (sandwiched between two ground planes) configurations, vias between the adjacent layers and composite vias connecting distant layers in the stack, chains of vias, etc. We also studied the effect of moats − long and narrow slits in the ground planes − on inductance of overlapping or adjacent microsrips and striplines. The presented set of data should be sufficient for designing circuits into fabrication processes for superconductor electronics developed at MIT Lincoln Laboratory such as SFQ5ee and SC1/SC2 processes. We especially focused on inductors with deep submicron dimensions, down to 120 nm, in order to provide input for circuit design into the most advanced process nodes, in particular, the newest 150-nm node of the SC2 process.

In majority of cases we observed good agreement between the existing inductance extractors, wxLL and InductEx, and the measured values without using any fitting parameters if the actually measured linewidth and dielectric thicknesses are used in the layer definition files of the software. However, agreement for more complex 3D objects such as vias is not so good and more development is required, especially to describe vias of very small sizes and stacked-staggered vias possessing a very significant inductance.

Moats in ground planes are essential components of integrated circuits, allowing to trap and keep magnetic flux outside of flux-bias-sensitive parts of the circuit. Strong coupling between circuit inductors and flux trapped in the moats can occur if the circuit is cooled in a nonuniform magnetic field changing sign (direction) within the circuits. Similar coupling can be induced upon magnetic filed cooling if the field is switched off in the superconducting state.

At small linewidths of and spacings between inductors, which are required for increasing integration scale of superconductor electronics, mutual inductance becomes significant and may induce unwanted parasitic coupling. Mutual inductance is important for designing meandered inductors and transformers. Mutual inductance will be covered in part II of the paper.

## Acknowledgment

We are grateful to Vasili Semenov for numerous discussions of inductance measurements and extraction as well as of flux trapping in superconductor integrated circuits. S.K. Tolpygo would like to thank Mikhail M. Khapaev for the excess to inductance extraction software wxLL and wxLC, and to Coenrad J. Fourie for the access to and help with InductEx. The authors are grateful to Alex Wynn for developing and maintaining a database of inductance measurements, to Ravi Rastogi and Scott Zarr for their help with wafer fabrication and to the entire MIT LL fabrication team. We would like to thank Leonard Johnson, Mark Gouker, Scott Holmes, and Mark Heiligman for their interest in and support of this work.

This research was based upon work supported by the Office of the Director of National Intelligence (ODNI), Intelligence Advanced Research Projects Activity (IARPA), via Air Force Contract FA8702-15-D-0001. The views and conclusions contained in this publication are those of the authors and should not be interpreted as necessarily representing the official policies or endorsements, either expressed or implied, of the ODNI, IARPA, or the U.S. Government. The U.S. Government is authorized to reproduce and distribute reprints for Governmental purposes notwithstanding any copyright annotation thereon.